\documentclass[11pt]{article}

\usepackage[margin=1in]{geometry}
\usepackage[T1]{fontenc}
\usepackage[utf8]{inputenc}
\usepackage{lmodern}
\usepackage{microtype}
\usepackage{amsmath,amssymb}
\usepackage{booktabs}
\usepackage{tabularx}
\usepackage{multirow}
\usepackage{array}
\usepackage{graphicx}
\usepackage{xcolor}
\usepackage{tikz}
\usetikzlibrary{arrows.meta,positioning,shapes.geometric,fit,calc}
\usepackage{listings}
\usepackage{enumitem}
\usepackage{cite}
\usepackage{caption}
\usepackage[most]{tcolorbox}
\usepackage[hidelinks]{hyperref}
\usepackage[capitalize,noabbrev]{cleveref}

\definecolor{AccentBlue}{HTML}{2A78D6}
\definecolor{AccentGreen}{HTML}{1BAF7A}
\definecolor{AccentRed}{HTML}{EB6834}
\definecolor{BoxBg}{HTML}{F7F9FB}
\definecolor{BoxBorder}{HTML}{D0D8E2}
\definecolor{CodeKw}{HTML}{005CC5}
\definecolor{CodeCmt}{HTML}{5C6370}
\definecolor{InkMuted}{HTML}{52514E}

\lstdefinelanguage{VerilogCustom}{
  keywords={module,endmodule,input,output,wire,reg,parameter,localparam,always,begin,end,if,else,case,endcase,default,assign,posedge,negedge,for,always_comb,generate,genvar,endgenerate,logic,initial},
  comment=[l]{//},
  morecomment=[s]{/*}{*/},
  morestring=[b]",
  sensitive=true
}
\newcommand{\code}[1]{\texttt{#1}}
\newcommand{\pp}{\,\mathrm{pp}}
\newcommand{\bench}{\textsc{RTLReviewBench}}

\newtcolorbox{answerbox}[1]{enhanced, breakable, colback=BoxBg, colframe=BoxBorder,
  boxrule=0.6pt, arc=2pt, left=6pt, right=6pt, top=4pt, bottom=4pt,
  fontupper=\small, before skip=6pt, after skip=8pt,
  title={\small\bfseries Answer to #1}, coltitle=black, colbacktitle=BoxBorder!45,
  attach title to upper={\ }, }

\title{\bfseries RTLReviewBench: LLM Hardware Reviewers\\ Rarely Miss Bugs but Routinely Reject Correct RTL}
\author{Yasir Mahmud\\University of Florida\\\texttt{ymahmud@ufl.edu}}
\date{}

\begin{document}
\maketitle

\begin{abstract}
Hardware teams increasingly ask large language models (LLMs) to review register-transfer level (RTL) code, yet it is unclear whether a model can tell a correct module from a buggy one without simulating it. We introduce \bench{}, a benchmark of 114 matched pairs of buggy and correct Verilog modules. Each pair shares one specification, most buggy modules are real failures of LLM code generators, and every label is confirmed by simulation. Evaluating 12 commercial and open-weight models, we find a consistent asymmetry: reviewers rarely accept buggy code but often reject correct code, producing more than five false rejections for every missed bug. Synthetic mutants hide part of the problem: on the same designs, models miss far fewer injected bugs than real ones. Asking a model to reason before it commits to a verdict sharply reduces false rejections for weaker reviewers, and a controlled ablation shows that \emph{where} the reasoning appears in the output matters more than the instruction to reason, whereas sampling and majority voting do not help. A blinded audit further shows that some correct ``buggy'' verdicts rest on false technical claims. We release the benchmark, a harder split for future models, all model responses, and the evaluation code.
\end{abstract}

\section{Introduction}
\label{sec:intro}

LLMs now write a large share of first-draft code, and hardware design is no exception: generation benchmarks for Verilog report steadily rising pass rates~\cite{liu2023verilogeval,pinckney2025verilogeval,lu2024rtllm,pinckney2025cvdp}. A natural next step is to use the same models as \emph{reviewers}, checking a frozen RTL module against its specification before it enters an expensive simulation or formal sign-off loop. Review is attractive because it is cheap and needs no testbench, but it is only useful if the reviewer is reliable in both directions. A reviewer that accepts buggy RTL lets defects drift toward silicon; a reviewer that rejects correct RTL floods engineers with false alarms until its warnings are ignored.

Existing benchmarks do not measure this skill. RTL generation suites score the code a model writes, not its judgment of code it is given. Debugging and repair suites~\cite{tsai2024rtlfixer,meng2026veribugbench,yu2026chipbench,zhu2026rtlrepairbench} reward a passing patch, which a model can often produce by regenerating the module from the specification without understanding the original fault~\cite{zhu2026rtlrepairbench}. Software code-review and LLM-as-a-judge benchmarks~\cite{jiang2025codejudgebench,zeng2026swrbench,jin2026overcorrection} do separate the two error directions, and Jin and Chen~\cite{jin2026overcorrection} show that LLMs over-reject correct Python, but these benchmarks contain none of the semantics that make hardware review hard: clocked state, blocking versus non-blocking assignment, reset style, and context-dependent bit widths.

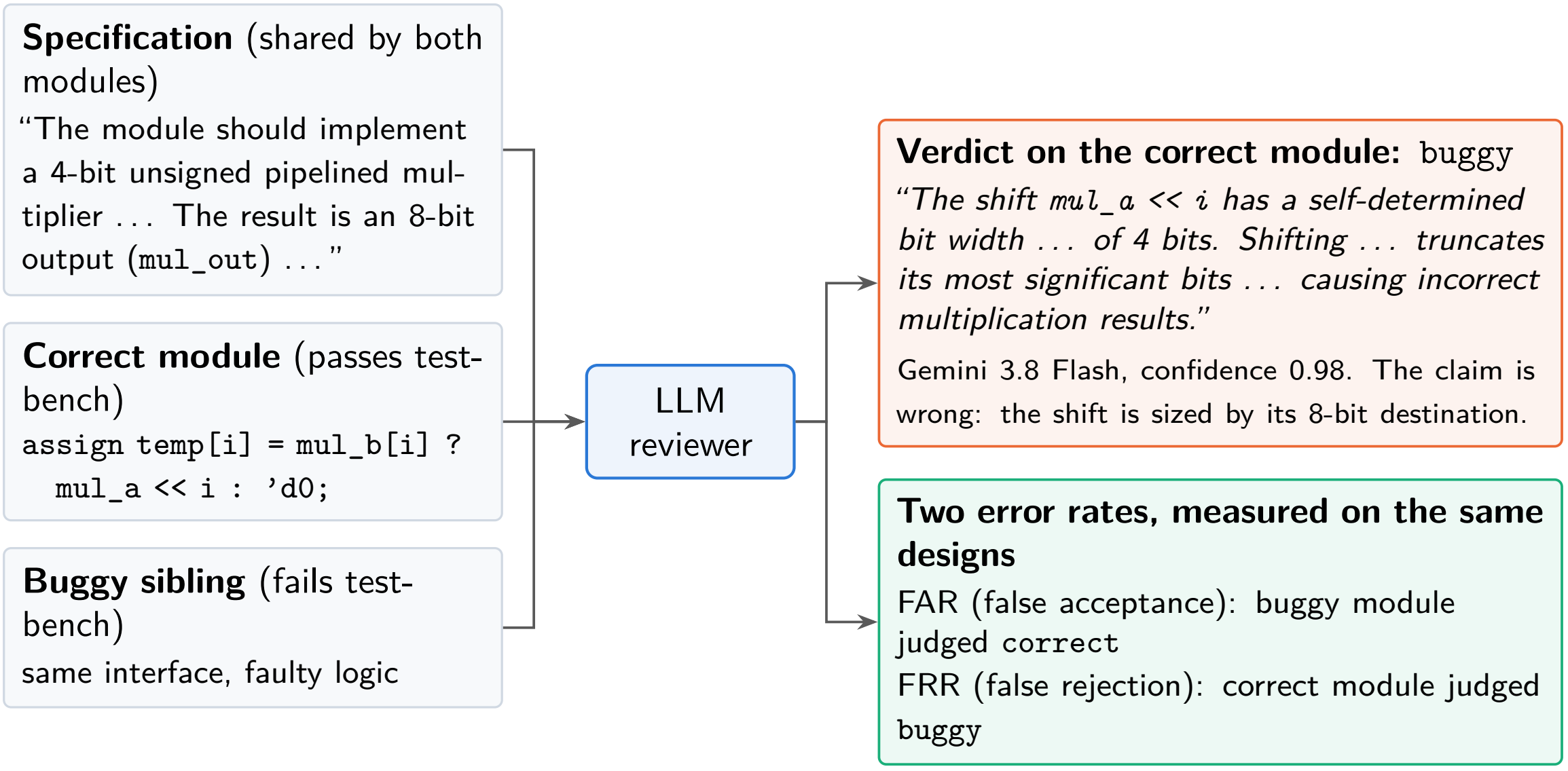
\begin{figure}[t]
\centering
\begin{tikzpicture}[
  font=\small\sffamily,
  doc/.style={rectangle, rounded corners=2pt, draw=BoxBorder, fill=BoxBg, line width=0.6pt, align=left, inner sep=5pt, text width=4.5cm},
  model/.style={rectangle, rounded corners=3pt, draw=AccentBlue, fill=AccentBlue!8, line width=0.8pt, align=center, inner sep=6pt, minimum height=1.1cm, text width=1.6cm},
  out/.style={rectangle, rounded corners=2pt, draw=#1, fill=#1!8, line width=0.7pt, align=left, inner sep=5pt, text width=6.3cm},
  arr/.style={-{Stealth[length=2mm]}, line width=0.7pt, draw=black!65}]
\node[doc] (spec) {\textbf{Specification} (shared by both modules)\\[1pt]\footnotesize ``The module should implement a 4-bit unsigned pipelined multiplier \dots\ The result is an 8-bit output (\texttt{mul\_out}) \dots''};
\node[doc, below=0.25cm of spec] (cor) {\textbf{Correct module} (passes testbench)\\[1pt]\footnotesize\ttfamily assign temp[i] = mul\_b[i] ?\\\quad mul\_a <{}< i : 'd0;};
\node[doc, below=0.25cm of cor] (bug) {\textbf{Buggy sibling} (fails testbench)\\[1pt]\footnotesize same interface, faulty logic};
\node[model, right=0.8cm of cor] (rev) {LLM\\reviewer};
\node[out=AccentRed, right=0.8cm of rev, yshift=1.35cm] (fr) {\textbf{Verdict on the correct module:} \code{buggy}\\[2pt]\footnotesize\itshape ``The shift \texttt{mul\_a <{}< i} has a self-determined bit width \dots\ of 4 bits. Shifting \dots\ truncates its most significant bits \dots\ causing incorrect multiplication results.''\\[2pt]\upshape\scriptsize Gemini 3.8 Flash, confidence 0.98. The claim is wrong: the shift is sized by its 8-bit destination.};
\node[out=AccentGreen, below=0.3cm of fr] (ok) {\textbf{Two error rates, measured on the same designs}\\[2pt]\footnotesize FAR (false acceptance): buggy module judged \code{correct}\\FRR (false rejection): correct module judged \code{buggy}};
\draw[arr] (spec.east) -- ++(0.3,0) |- (rev.west);
\draw[arr] (cor.east) -- (rev.west);
\draw[arr] (bug.east) -- ++(0.3,0) |- (rev.west);
\draw[arr] (rev.east) -- ++(0.3,0) |- (fr.west);
\draw[arr] (rev.east) -- ++(0.3,0) |- (ok.west);
\end{tikzpicture}
\caption{The \bench{} task. A reviewer sees the specification and \emph{one} of two sibling modules, never both, and must judge whether the module conforms. The example is a real false rejection caused by a misapplied Verilog width rule.}
\label{fig:task}
\end{figure}

We introduce \bench{}, a benchmark for execution-free RTL review (\cref{fig:task}). Each item pairs a specification with two modules that compile and share the same ports: a reference that passes the upstream testbench, and a buggy module that fails it. Most buggy modules (97 of 114) are real functional failures of LLM code generators from \textsc{RTLRepairBench}~\cite{zhu2026rtlrepairbench}; the rest are hand-written faults from \textsc{ChipBench}~\cite{yu2026chipbench}. None are produced by automatic mutation. A reviewer sees one module at a time and must return \code{correct} or \code{buggy}. Because the two siblings share a specification, we can measure missed bugs (false acceptance) and false alarms (false rejection) on the same designs.

Using \bench{}, we evaluate 12 models from four providers and ask four questions:

\begin{itemize}[leftmargin=1.4em,itemsep=1pt]
  \item \textbf{RQ1:} How reliably can LLMs review RTL, and which way do their errors go?
  \item \textbf{RQ2:} Do synthetic mutants, the standard way to build bug benchmarks, measure the same thing as real bugs?
  \item \textbf{RQ3:} Can prompting, sampling, or side-by-side comparison reduce review errors?
  \item \textbf{RQ4:} When a reviewer flags a buggy module, is its explanation correct?
\end{itemize}

Our main findings are that (i)~errors are strongly asymmetric: 11 of the 12 models miss at most 3.5\% of bugs, while false rejection ranges from 2\% to 59\% of correct modules; (ii)~synthetic mutants make reviewers look safer than they are on the same designs; (iii)~putting the reasoning before the verdict in the output schema removes much of the over-rejection of weaker models, while majority voting over samples does not; and (iv)~a fraction of correct ``buggy'' verdicts cite defects that do not exist.

\paragraph{Contributions.}
\begin{enumerate}[leftmargin=1.4em,itemsep=1pt]
  \item \textbf{A benchmark of simulation-verified RTL bugs in matched pairs, mostly real LLM failures} (\cref{sec:benchmark}), with an audit that removes six upstream items whose testbench failures stem from simulator artifacts rather than logic errors, a matched synthetic-mutant set, and a held-out \textbf{Hard} split on which the best current reviewer gets both siblings right on only 77\% of pairs.
  \item \textbf{An evaluation of 12 reviewers} (\cref{sec:rq1,sec:rq2}) showing that false rejection, not missed bugs, is the dominant failure, and that synthetic mutants understate how often real bugs are missed.
  \item \textbf{Controlled studies of what helps} (\cref{sec:rq3,sec:rq4}): a $2{\times}2$ ablation that separates output ordering from the instruction to reason, comparisons with self-consistency and pairwise review, and a blinded audit of reviewer explanations.
\end{enumerate}
All data, 17{,}623 model and judge responses, and code are released.\footnote{\url{https://github.com/yasirmahmud/RTLReviewBench}}

\section{Related Work}
\label{sec:related}

\paragraph{RTL generation.} \textsc{VerilogEval}~\cite{liu2023verilogeval,pinckney2025verilogeval}, \textsc{RTLLM}~\cite{lu2024rtllm}, and \textsc{VeriGen}~\cite{thakur2024verigen} established simulation-based pass@$k$ for single modules; \textsc{CVDP}~\cite{pinckney2025cvdp}, \textsc{RealBench}~\cite{jin2025realbench}, \textsc{TuRTLe}~\cite{garciagasulla2025turtle}, and \textsc{RTL-BenchLS}~\cite{wang2026rtlbenchls} extend coverage to verification tasks, formal equivalence, and larger designs. All of them grade code the model writes. \bench{} instead fixes the code and grades the model's judgment of it.

\paragraph{RTL debugging and repair.} \textsc{RTLFixer}~\cite{tsai2024rtlfixer} repairs syntax errors; \textsc{ChipBench}~\cite{yu2026chipbench}, \textsc{VeriBugBench}~\cite{meng2026veribugbench}, \textsc{HWE-Bench}~\cite{cui2026hwebench}, and \textsc{Phoenix-bench}~\cite{zou2026phoenixbench} evaluate functional repair from single modules up to whole repositories. \textsc{RTLRepairBench}~\cite{zhu2026rtlrepairbench} shows that repair scores can be earned by regenerating code from the specification and that simulation oracles carry reset and time-zero confounds. We reuse its real LLM failures and adopt its concern with oracle validity, but ask for a verdict rather than a patch, which removes the regeneration shortcut.

\paragraph{Code review and LLM-as-a-judge.} \textsc{CodeJudgeBench}~\cite{jiang2025codejudgebench} finds that judges are sensitive to response order; \textsc{SWR-Bench}~\cite{zeng2026swrbench} and \textsc{MCR-Bench}~\cite{zheng2026mcrbench} evaluate review comments on pull requests; and Jin and Chen~\cite{jin2026overcorrection} report that LLMs systematically reject correct code and that richer prompts can make this worse. Related work documents position bias~\cite{zheng2023judging}, the effect of output-format constraints on reasoning~\cite{tam2024speak}, and gains from self-consistency~\cite{wang2023selfconsistency}. We test whether these findings carry over to hardware, where correctness depends on timing and bit-level semantics. \Cref{tab:related} summarizes the differences.

\begin{table}[t]
\centering
\caption{\bench{} compared with closely related benchmarks.}
\label{tab:related}
\small
\setlength{\tabcolsep}{4pt}
\begin{tabularx}{\textwidth}{@{}l l >{\raggedright\arraybackslash}X c c@{}}
\toprule
\textbf{Benchmark} & \textbf{Domain} & \textbf{Task} & \textbf{Matched pairs} & \textbf{FAR \& FRR} \\
\midrule
VerilogEval v2~\cite{pinckney2025verilogeval} & RTL & Generate RTL from a specification & -- & -- \\
ChipBench~\cite{yu2026chipbench} & RTL & Generate, debug, and write reference models & -- & -- \\
VeriBugBench~\cite{meng2026veribugbench} & RTL & Localize and repair injected faults & -- & -- \\
RTLRepairBench~\cite{zhu2026rtlrepairbench} & RTL & Repair real and synthetic faults & partial & -- \\
CodeJudgeBench~\cite{jiang2025codejudgebench} & Software & Rank two candidate responses & \checkmark & -- \\
SWR-Bench~\cite{zeng2026swrbench} & Software & Comment on pull requests & \checkmark & \checkmark \\
Jin \& Chen~\cite{jin2026overcorrection} & Software & Judge one implementation & \checkmark & \checkmark \\
\midrule
\textbf{\bench{} (ours)} & \textbf{RTL} & \textbf{Judge one frozen module, no repair} & \checkmark & \checkmark \\
\bottomrule
\end{tabularx}
\end{table}

\section{The \bench{} Benchmark}
\label{sec:benchmark}

\subsection{Task}
\label{sec:task}

An item is a design $k$ with specification $S_k$ and two modules $C_k^{\mathrm{cor}}$ and $C_k^{\mathrm{bug}}$ that compile and have identical interfaces. $C_k^{\mathrm{cor}}$ passes the upstream reference testbench and $C_k^{\mathrm{bug}}$ fails it. The reviewer receives $S_k$ and a single module, never both, together with the instruction ``Judge only the supplied RTL implementation. Do not write or repair RTL,'' and returns a JSON object with a verdict (\code{correct}, \code{buggy}, or \code{abstain}), a confidence, and a short reason. File names, comments that hint at the answer, and sibling information are removed. Exact prompts are in \cref{app:prompts}.

\subsection{Construction}
\label{sec:construction}

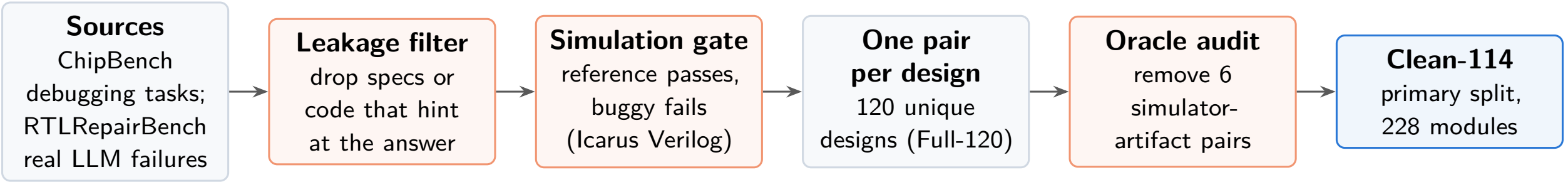
\begin{figure}[t]
\centering
\resizebox{\textwidth}{!}{%
\begin{tikzpicture}[
  font=\small\sffamily, node distance=0.5cm,
  stage/.style={rectangle, rounded corners=3pt, draw=BoxBorder, fill=BoxBg, line width=0.7pt, align=center, inner sep=5pt, minimum height=1.35cm, text width=2.55cm},
  gate/.style={stage, draw=AccentRed!70, fill=AccentRed!6},
  final/.style={stage, draw=AccentBlue, fill=AccentBlue!7},
  arr/.style={-{Stealth[length=2mm]}, line width=0.7pt, draw=black!65}]
\node[stage] (s1) {\textbf{Sources}\\\footnotesize ChipBench debugging tasks; RTLRepairBench real LLM failures};
\node[gate, right=of s1] (s2) {\textbf{Leakage filter}\\\footnotesize drop specs or code that hint at the answer};
\node[gate, right=of s2] (s3) {\textbf{Simulation gate}\\\footnotesize reference passes, buggy fails (Icarus Verilog)};
\node[stage, right=of s3] (s4) {\textbf{One pair per design}\\\footnotesize 120 unique designs (Full-120)};
\node[gate, right=of s4] (s5) {\textbf{Oracle audit}\\\footnotesize remove 6 simulator-artifact pairs};
\node[final, right=of s5] (s6) {\textbf{Clean-114}\\\footnotesize primary split, 228 modules};
\foreach \a/\b in {s1/s2,s2/s3,s3/s4,s4/s5,s5/s6} \draw[arr] (\a) -- (\b);
\end{tikzpicture}}
\caption{Construction of the primary split. The same filters produce the synthetic-mutant set (SynMut-50).}
\label{fig:pipeline}
\end{figure}

\Cref{fig:pipeline} summarizes the pipeline; full accounting is in \cref{app:construction}.

\paragraph{Sources.} We draw real bugs from two pinned corpora. \textsc{ChipBench}~\cite{yu2026chipbench} provides sequential debugging tasks in which engineers injected faults by hand into golden modules. \textsc{RTLRepairBench}~\cite{zhu2026rtlrepairbench} provides \code{RealBugs}: functional failures of Qwen2.5-Coder models on \textsc{VerilogEval}~v2 tasks, each paired with the upstream reference.

\paragraph{Filtering and verification.} We remove tasks whose specification is incomplete or contradicts the reference, and items whose specification or code leaks the answer (for example, a comment marking the correct line). Every surviving pair is compiled and simulated with Icarus Verilog~11.0~\cite{williams2002iverilog}; we keep a pair only if the reference passes its testbench and the buggy module fails it. To avoid counting the same design twice, we keep one pair per upstream task, giving 120 designs (\textbf{Full-120}).

\paragraph{Oracle audit.} A failing testbench does not always mean a logic bug. Inspecting low-mismatch failures, we found six pairs in which the ``buggy'' module is functionally equivalent to the reference and fails only because of simulator semantics: an explicit \code{\#delay}, a missing time-zero event, or a reference that relies on a non-synthesizable \code{initial} block (\cref{app:artifact6}). These six pairs (\textbf{Artifact-6}) account for half of all missed bugs on Full-120, so we report results on the remaining \textbf{Clean-114} split and treat Full-120 as a sensitivity check.

\paragraph{Additional splits.} \Cref{tab:splits} lists all splits. \textbf{SynMut-50} contains single-operator mutants of \textsc{VerilogEval} references from \textsc{RTLRepairBench}, passed through the same filters. On 42 of its designs, the same reference also appears in Clean-114 with a real LLM bug (\textbf{Matched-42}), which lets us compare real and synthetic bugs on identical designs. \textbf{Hard} contains the 22 Clean-114 pairs on which at least half of our reviewers err on a sibling; because it is selected with the evaluated models, we intend it as a target for future models rather than an unbiased estimate for these ones.

\begin{table}[t]
\centering
\caption{Benchmark splits. All splits contain one buggy and one correct module per design.}
\label{tab:splits}
\small
\begin{tabularx}{\textwidth}{@{}l c X@{}}
\toprule
\textbf{Split} & \textbf{Designs} & \textbf{Purpose} \\
\midrule
\textbf{Clean-114} & 114 & Primary evaluation: 97 real LLM failures (RTLRepairBench) and 17 hand-written faults (ChipBench); 80 sequential designs \\
Full-120 & 120 & Sensitivity check: Clean-114 plus the six Artifact-6 pairs \\
SynMut-50 & 50 & Single-operator synthetic mutants (six operator types) \\
Matched-42 & 42 & Designs present in both Clean-114 and SynMut-50 \\
Hard & 22 & Pairs that at least 6 of 12 reviewers get wrong; headroom for future models \\
\bottomrule
\end{tabularx}
\end{table}

\section{Experimental Setup}
\label{sec:setup}

\paragraph{Models.} We evaluate 12 models from four providers (\cref{tab:models}): Claude Haiku 5.5 (with and without extended thinking) and Claude Sonnet 4.6 from Anthropic; Gemini 3.1 Pro, Gemini 3.8 Flash, and the open-weight Gemma 4 31B from Google; GPT-5.4, GPT-4.1, GPT-4.1 mini, o4-mini, and the open-weight gpt-oss-120b from OpenAI; and Llama 3.3 70B from Meta. Five of them spend hidden reasoning tokens before answering. All runs use greedy decoding ($T=0$), JSON output mode, no examples, and no tools; queries were issued through an OpenAI-compatible endpoint in a single two-day window (\cref{app:telemetry}).

\paragraph{Metrics.} We treat \code{buggy} as the positive class. The \emph{false acceptance rate} (FAR) is the fraction of buggy modules judged \code{correct}, and the \emph{false rejection rate} (FRR) is the fraction of correct modules judged \code{buggy}. Accuracy counts abstentions and invalid outputs as wrong. Because siblings share a specification, we also report \emph{pair accuracy}, the fraction of designs for which both siblings are classified correctly. We give exact 95\% Clopper--Pearson intervals~\cite{clopper1934use}, compare conditions on the same items with exact McNemar tests~\cite{mcnemar1947note}, and correct each family of tests with the Holm--Bonferroni procedure~\cite{holm1979simple}.

\section{Results}
\label{sec:results}

\subsection{RQ1: How Reliable Is LLM Review of RTL?}
\label{sec:rq1}

\begin{figure}[t]
\centering
\includegraphics[width=\textwidth]{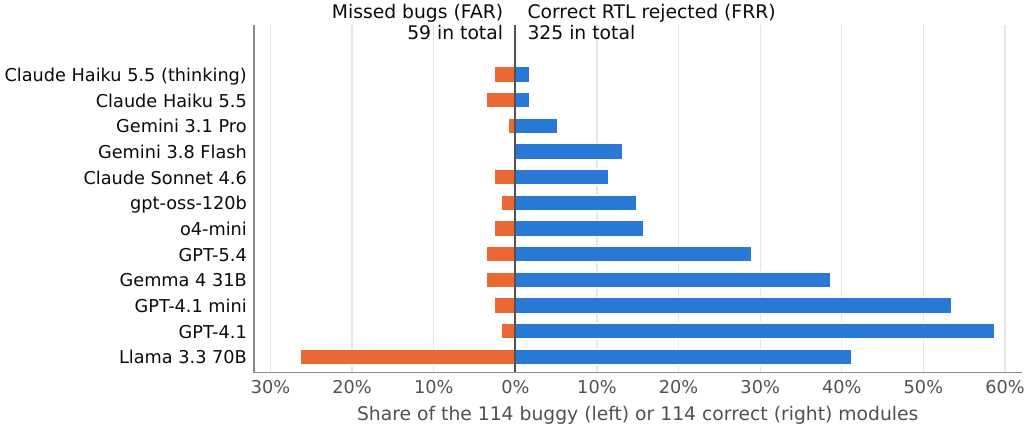}
\caption{Error direction on Clean-114. Bars to the left show the share of buggy modules accepted (FAR); bars to the right show the share of correct modules rejected (FRR). Across all models, correct modules are rejected 325 times and bugs are missed 59 times.}
\label{fig:asymmetry}
\end{figure}

\begin{table}[t]
\centering
\caption{Clean-114 results (228 modules per model, sorted by accuracy). Brackets give 95\% confidence intervals. Pair: both siblings correct. The full table, with Full-120 results, calibration, and latency, is \cref{tab:overall}.}
\label{tab:main}
\small
\setlength{\tabcolsep}{6pt}
\begin{tabular}{@{}l c c c c@{}}
\toprule
\textbf{Reviewer} & \textbf{Accuracy} & \textbf{FAR} & \textbf{FRR} & \textbf{Pair} \\
\midrule
Claude Haiku 5.5 (thinking) & \textbf{97.4} {\scriptsize[94.4, 99.0]} & 2.6 & \textbf{1.8} & \textbf{94.7} \\
Claude Haiku 5.5 & \textbf{97.4} {\scriptsize[94.4, 99.0]} & 3.5 & \textbf{1.8} & \textbf{94.7} \\
Gemini 3.1 Pro & 96.9 {\scriptsize[93.8, 98.8]} & 0.9 & 5.3 & 93.9 \\
Gemini 3.8 Flash & 93.4 {\scriptsize[89.4, 96.3]} & \textbf{0.0} & 13.2 & 86.8 \\
Claude Sonnet 4.6 & 93.0 {\scriptsize[88.9, 95.9]} & 2.6 & 11.4 & 87.7 \\
gpt-oss-120b & 91.7 {\scriptsize[87.3, 94.9]} & 1.8 & 14.9 & 84.2 \\
o4-mini & 90.8 {\scriptsize[86.3, 94.2]} & 2.6 & 15.8 & 82.5 \\
GPT-5.4 & 83.8 {\scriptsize[78.3, 88.3]} & 3.5 & 28.9 & 69.3 \\
Gemma 4 31B & 78.9 {\scriptsize[73.1, 84.1]} & 3.5 & 38.6 & 57.9 \\
GPT-4.1 mini & 71.9 {\scriptsize[65.6, 77.7]} & 2.6 & 53.5 & 45.6 \\
GPT-4.1 & 69.7 {\scriptsize[63.3, 75.6]} & 1.8 & 58.8 & 40.4 \\
Llama 3.3 70B & 66.2 {\scriptsize[59.7, 72.3]} & 26.3 & 41.2 & 36.0 \\
\midrule
Majority of three (cross-validated) & 97.4 & -- & -- & -- \\
\bottomrule
\end{tabular}
\end{table}

\Cref{tab:main} and \cref{fig:asymmetry} summarize single-module review on Clean-114.

\paragraph{The best reviewers are strong.} Claude Haiku 5.5 and Gemini 3.1 Pro classify about 97\% of modules correctly and get both siblings right on 94\% of designs. Gemini 3.8 Flash misses no bug at all, but pays for it by rejecting 13\% of correct modules. Larger or reasoning-enabled models do not reliably win: Claude Sonnet 4.6 rejects more correct modules than Claude Haiku 5.5, and turning on extended thinking for Haiku leaves its accuracy unchanged.

\paragraph{Errors are lopsided.} Eleven of the twelve models miss at most 3.5\% of bugs. What separates them is false rejection, which ranges from 2\% to 59\%. Pooled over all models, correct modules are rejected 5.5 times as often as bugs are missed (\cref{fig:asymmetry}). The weakest reviewers reject more than half of all correct modules, so in practice they would flag almost everything. Llama 3.3 70B is the only model that also misses many bugs.

\paragraph{Why correct code gets rejected.} Reading the rejections (\cref{app:cases}), we find three recurring causes: misapplied Verilog semantics, such as the width rule in \cref{fig:task}; invented requirements about latency, reset style, or coding conventions that the specification does not state; and dialect purism, such as rejecting legal SystemVerilog constructs. Longer, multi-block sequential designs from \textsc{ChipBench} trigger the most rejections. Only one module defeats all 12 reviewers, and it reveals a mismatch between an upstream specification and its reference rather than a reviewer failure (\cref{app:cases}).

\paragraph{Combining reviewers.} A majority vote over three models from different providers matches the best single model (97.4\%) when the triple is chosen by five-fold cross-validation; the best triple chosen in hindsight reaches 98.2\% (\cref{app:full}). At this level, ensembles offer at most a small gain.

\paragraph{Headroom.} On the Hard split, the best reviewer gets both siblings right on 77\% of pairs and mean accuracy drops from 86\% to 67\% (\cref{tab:hard}); false rejection still dominates.

\begin{answerbox}{RQ1}
The best LLMs review RTL accurately, but errors across models are strongly asymmetric: reviewers rarely accept a buggy module and often reject a correct one. For most models, false rejection, not missed bugs, is the main obstacle to deployment, so FAR and FRR should always be reported separately.
\end{answerbox}

\subsection{RQ2: Do Synthetic Mutants Reflect Real Bugs?}
\label{sec:rq2}

\begin{figure}[t]
\centering
\includegraphics[width=\textwidth]{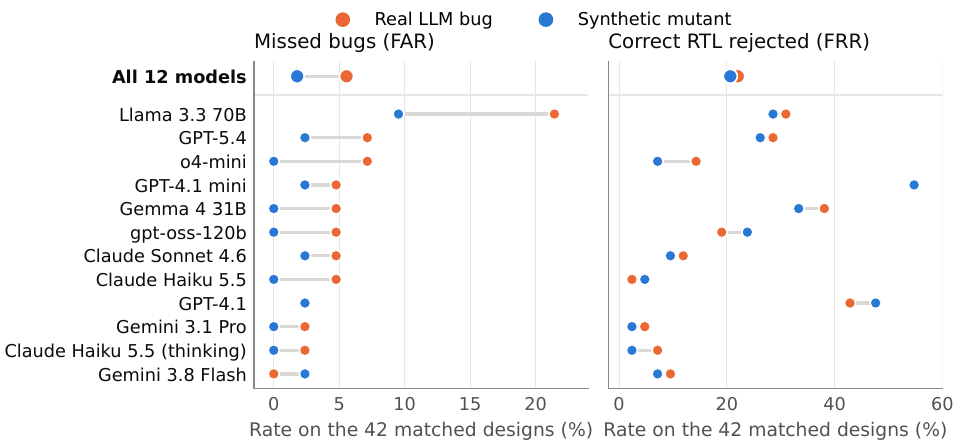}
\caption{Real LLM bugs versus synthetic mutants on the same 42 designs (Matched-42). Each row is a model; the top row pools all twelve. Synthetic mutants are missed less often (left) while false rejection of the shared correct reference barely changes (right).}
\label{fig:synthetic}
\end{figure}

Bug benchmarks are usually built by injecting single-operator mutations into correct code~\cite{meng2026veribugbench,zhu2026rtlrepairbench}. Matched-42 lets us ask whether such mutants are as hard to catch as real bugs, holding the design and the correct reference fixed.

\paragraph{Mutants are easier to catch.} Pooled over all twelve models, reviewers miss 28 of 504 real bugs but only 9 of 504 synthetic mutants on the same designs (exact McNemar $p<0.001$; \cref{fig:synthetic}). The direction holds for ten of the twelve models. One plausible explanation is that a mutant such as a swapped operator or a flipped comparison changes a single visible token, whereas an LLM-written bug is embedded in code that was generated to look right.

\paragraph{False rejection is unchanged.} The correct reference is identical in both conditions, and its rejection rate barely moves (22\% versus 21\%). Synthetic benchmarks therefore understate the risk of missed bugs while saying nothing new about false alarms.

\paragraph{Oracle artifacts distort the picture further.} Before the oracle audit, bugs produced by the fine-tuned generator appeared to be a blind spot, with most models missing 30--50\% of them. All of these apparent misses came from the three \code{\#delay} artifacts; on the remaining clean items, nine of twelve models miss none.

\begin{answerbox}{RQ2}
No. On identical designs, synthetic mutants are missed far less often than real LLM bugs, and they leave the dominant failure, false rejection, untouched. Benchmarks for RTL review should use real bugs and audit their simulation oracles.
\end{answerbox}

\subsection{RQ3: Can Prompting, Sampling, or Comparison Help?}
\label{sec:rq3}

\begin{figure}[t]
\centering
\includegraphics[width=\textwidth]{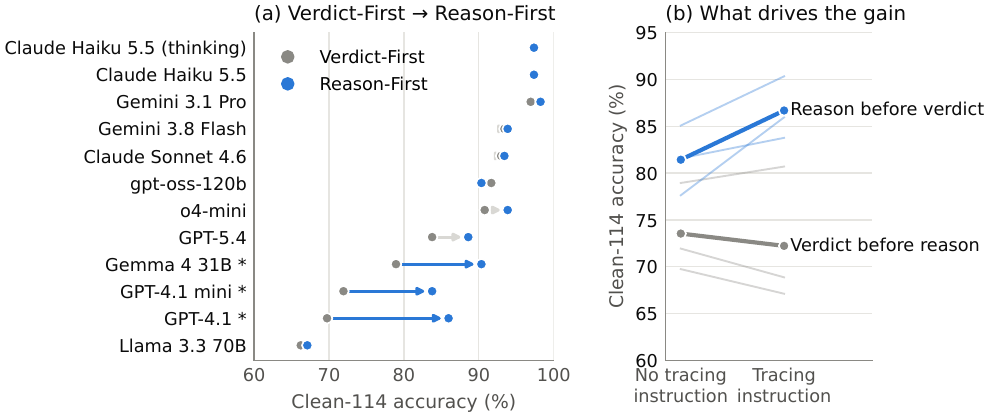}
\caption{\textbf{(a)} Accuracy with the default Verdict-First schema (grey) and with Reason-First prompting (blue); * marks Holm-significant changes. \textbf{(b)} A $2{\times}2$ ablation on the three models marked * separates the order of fields in the output from the instruction to trace the logic. Thick lines are means; thin lines are individual models.}
\label{fig:reason_first}
\end{figure}

We test three common ways to improve a judge: changing the prompt, sampling several answers, and comparing candidates side by side.

\paragraph{Reason-First prompting.} Our default schema asks for the verdict first and the reason afterwards. \emph{Reason-First} prompting makes two changes: the \code{reason} field comes before \code{verdict}, and the reviewer is told to trace signals, operators, and timing step by step before deciding. As \cref{fig:reason_first}(a) shows, this improves three short-answer models by 11 to 16 points, mostly by cutting false rejection (GPT-4.1 goes from 70\% to 86\%). Models that already reason at length change little, although Gemini 3.1 Pro reaches 98.2\%, the best single result in the study. Reason-First also removes a telling failure of the default schema: in 15 Clean responses, a model emitted a verdict, worked through the logic, and concluded the opposite inside its own explanation. For example, Gemma 4 answered \code{"verdict":"correct"} for an XNOR module and then wrote \emph{``This is XOR. The spec asks for XNOR. The implementation is buggy.''} With the reason first, all 15 are resolved correctly.

\paragraph{Order matters more than the instruction.} To separate the two changes, we ran the two missing cells of a $2{\times}2$ design on the three models that benefited (\cref{fig:reason_first}(b)). Moving the reason before the verdict, without any tracing instruction, raises accuracy by 8 points on average and is significant for all three models. Adding the tracing instruction while the verdict still comes first triples output length but does not help (a 1-point drop on average). The full gain requires both: tracing helps only when it happens before the decision, a positive interaction of about 7 points (95\% bootstrap interval 3.5 to 9.6). The output schema thus acts as a commitment device. Once a model has emitted a verdict, later reasoning rarely changes its decision.

\paragraph{Sampling and voting do not help.} For three commercial models, we drew five samples at $T=0.7$ and took the majority. Accuracy changed by between $-1.3$ and $0$ points, because 93--94\% of items received five identical votes: the errors are systematic, not random. Voting across different models (RQ1) helps more than voting within one.

\paragraph{Side-by-side comparison helps but adds position bias.} When shown both siblings and asked which one conforms, most models pick the correct module more reliably than they classify the two modules separately; Gemini 3.1 Pro picks the correct module under both orderings on 98\% of pairs. Weaker models, however, often change their answer when the two candidates are swapped, mostly in favour of whichever module is shown first; Llama 3.3 70B does so on 20 of 114 pairs, echoing position bias in software judges~\cite{zheng2023judging,jiang2025codejudgebench}. Comparison is also not available in real review, where no correct sibling exists. Full results for all three interventions are in \cref{app:full}.

\begin{answerbox}{RQ3}
Prompting helps when it changes \emph{when} the model reasons: placing the reason before the verdict recovers most of the over-rejection of weaker reviewers, while an instruction to reason after committing does not help. Self-consistency sampling does not fix systematic errors, and side-by-side comparison introduces position bias in weaker models.
\end{answerbox}

\subsection{RQ4: Are Correct Verdicts Correctly Explained?}
\label{sec:rq4}

\begin{figure}[t]
\centering
\includegraphics[width=\textwidth]{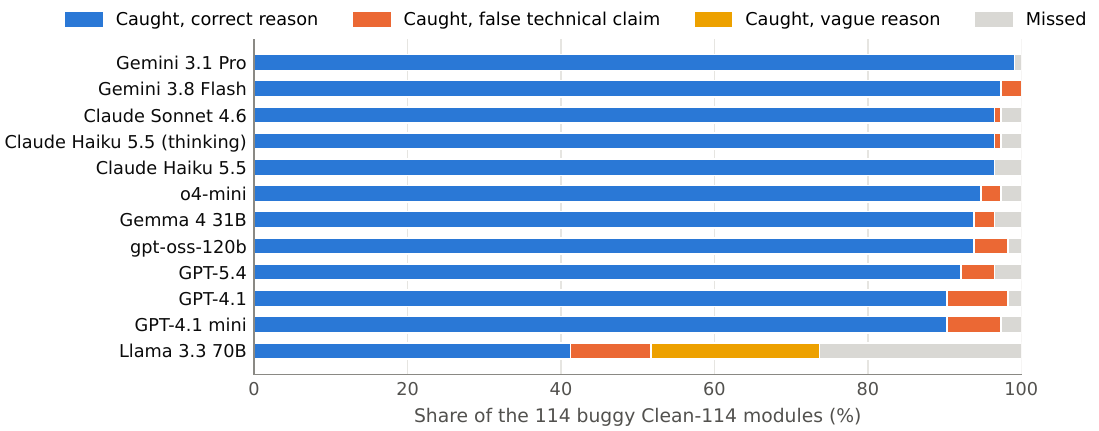}
\caption{Explanation audit of every buggy module flagged on Clean-114. A detection counts as ``correct reason'' only if the explanation identifies the actual defect.}
\label{fig:rationale}
\end{figure}

A verdict of \code{buggy} on a buggy module counts as correct even if the stated reason is wrong. A reviewer that rejects most code will also ``catch'' most bugs, often for the wrong reason. We therefore audited all 1{,}309 correct detections on Clean-114. Two judges from different providers (GPT-5.4 and Claude Sonnet 4.6) saw the specification, both siblings, and the failing simulation output, but not the reviewer's identity, and labelled each explanation \emph{grounded} (names the real defect), \emph{spurious} (asserts a defect that does not exist), or \emph{vague}. A third model (Gemini 3.1 Pro) resolved disagreements, and a human checked every item labelled spurious.

\paragraph{Most detections are well founded, but not all.} Pooled across models, 94\% of detections cite the real defect, and the strongest reviewers are almost always right for the right reason (\cref{fig:rationale}). The over-rejecting models lose the most: GPT-4.1 and GPT-4.1 mini lose eight to nine detections each to false technical claims. Llama 3.3 70B correctly explains only 41\% of buggy modules, although it labels 74\% of them buggy. In one case, GPT-4.1 and GPT-5.4 both flagged a module with an incorrect asynchronous reset, but cited a legal SystemVerilog construct as the defect and never mentioned the reset.

\paragraph{Judges need checking too.} The two judges agree on 93\% of items (Cohen's $\kappa=0.54$~\cite{cohen1960coefficient}). Both rate their own family's explanations more favourably, by up to 9 points for Claude Sonnet 4.6, which is why we use cross-provider judges, a tie-breaker, and human review. The human check confirmed 48 of the 50 spurious labels.

\paragraph{Confidence tracks reasoning effort.} Models that answer in under 100 tokens are also badly calibrated: they report near-certain confidence while being wrong a sixth to a third of the time (Gemma 4 reports full confidence on 224 of 228 modules). Models that reason at length are well calibrated (\cref{tab:overall}).

\begin{answerbox}{RQ4}
Mostly, for strong reviewers. For weaker reviewers, a correct verdict is often a by-product of rejecting everything, and some cite defects that do not exist. Evaluations of LLM reviewers should check explanations, not only verdicts.
\end{answerbox}

\section{Discussion}
\label{sec:discussion}

\paragraph{Implications for practice.} Our results suggest four practices for teams that use LLM reviewers. First, measure false rejection explicitly: it is the failure most likely to make a reviewer unusable. Second, request the reasoning before the verdict; this simple change to the output schema was the largest improvement we observed. Third, prefer combining different models, or deferring disagreements to simulation, over resampling one model. Fourth, treat explanations as claims to verify, especially from reviewers that reject often.

\paragraph{Implications for benchmark design.} Synthetic mutants and unaudited simulation oracles both distorted our measurements: the former hid missed bugs, and the latter manufactured an apparent blind spot. RTL benchmarks should use real defects where possible, report how oracle artifacts were handled, and evaluate both error directions on matched items.

\paragraph{Threats to validity.} \emph{Internal.} Labels come from upstream testbenches; we audited low-mismatch failures by hand, but a formal equivalence check would cover inputs that the testbenches do not. The rationale audit relies on LLM judges, which we mitigated with cross-provider judges and human review. \emph{Construct.} The task presents modules in isolation, which is stricter than review with access to a simulator or a diff. \emph{External.} Designs are single modules of 6--134 lines, and 97 of the 114 come from \textsc{VerilogEval}, whose references may appear in training data; models score higher on these than on the newer \textsc{ChipBench} items, although the two sources also differ in design size. Results may change as hosted models are updated; all responses are released so the numbers can be rechecked.

\section{Conclusion}
\label{sec:conclusion}

We introduced \bench{}, a benchmark of simulation-verified RTL bugs in matched pairs, most of them real LLM failures. Across twelve models, LLM reviewers rarely miss bugs but routinely reject correct RTL; synthetic mutants make them look safer than they are; reasoning before the verdict, rather than after it, removes much of the over-rejection; and some correct verdicts rest on false claims. We hope \bench{} and its Hard split help make LLM review a dependable step in hardware verification rather than a source of false alarms.

\bibliographystyle{plain}
\bibliography{refs}

\clearpage
\appendix
\sloppy

\section{Prompts}
\label{app:prompts}

\begin{lstlisting}[language={},basicstyle=\fontsize{7.2}{8.3}\selectfont\ttfamily,aboveskip=3pt,belowskip=3pt,caption={Frozen single-candidate (\code{Verdict-First} Baseline), \code{Reason-First} prompting, $2{\times}2$ ablation cells (\code{vf\_trace}, \code{rf\_order\_only}), \code{Pairwise} comparison, and blinded \code{Rationale Audit Judge} templates.},label={lst:prompts_all}]
=== 1. SINGLE-CANDIDATE VERDICT-FIRST (Core & SynMut-50 Baseline) ===
SYSTEM:
You are reviewing RTL for functional conformance to the given specification.
Judge only the supplied RTL implementation. Do not write or repair RTL.
If evidence is insufficient, you may abstain. Return only valid JSON matching
{"verdict":"correct|buggy|abstain","confidence":0.0,"reason":"short explanation"}.
Use one of the three literal verdict strings. Confidence must be between 0 and 1.
Do not assume that syntactically valid code is functionally correct.

USER:
SPECIFICATION (verbatim from the existing dataset):
{specification}

RTL IMPLEMENTATION (candidate under review):
```verilog
{candidate_rtl}
```

Is this RTL functionally consistent with the specification?
Output JSON only, with verdict, confidence, and a short reason.

=== 2. REASON-FIRST PROMPTING (Schema Order + Tracing Instruction) ===
SYSTEM:
You are reviewing RTL for functional conformance to the given specification.
Judge only the supplied RTL implementation. Do not write or repair RTL.
If evidence is insufficient, you may abstain. Return only valid JSON matching
{"reason":"step-by-step functional check of signals, operators, and cycle timing against the spec","verdict":"correct|buggy|abstain","confidence":0.0}.
First trace the RTL behavior inside "reason", then commit to one of the three literal verdict strings.
Confidence must be between 0 and 1. Do not assume that syntactically valid code is functionally correct.

USER:
[Same SPECIFICATION & RTL IMPLEMENTATION block]
Is this RTL functionally consistent with the specification?
Output JSON only, reasoning first in "reason", then giving "verdict" and "confidence".

=== 2a. ABLATION CELL A: VF_TRACE (Verdict-First Order + Tracing Instruction) ===
SYSTEM:
You are reviewing RTL for functional conformance to the given specification.
Judge only the supplied RTL implementation. Do not write or repair RTL.
If evidence is insufficient, you may abstain. Return only valid JSON matching
{"verdict":"correct|buggy|abstain","confidence":0.0,"reason":"step-by-step functional check of signals, operators, and cycle timing against the spec"}.
In "reason", trace the RTL behavior step by step against the specification.
Use one of the three literal verdict strings. Confidence must be between 0 and 1.
Do not assume that syntactically valid code is functionally correct.

USER:
[Same SPECIFICATION & RTL IMPLEMENTATION block]
Is this RTL functionally consistent with the specification?
Output JSON only, with verdict, confidence, and a step-by-step reason.

=== 2b. ABLATION CELL B: RF_ORDER_ONLY (Reason-First Order Only) ===
SYSTEM:
You are reviewing RTL for functional conformance to the given specification.
Judge only the supplied RTL implementation. Do not write or repair RTL.
If evidence is insufficient, you may abstain. Return only valid JSON matching
{"reason":"short explanation","verdict":"correct|buggy|abstain","confidence":0.0}.
Use one of the three literal verdict strings. Confidence must be between 0 and 1.
Do not assume that syntactically valid code is functionally correct.

USER:
[Same SPECIFICATION & RTL IMPLEMENTATION block]
Is this RTL functionally consistent with the specification?
Output JSON only, with reason, verdict, and confidence, in that key order.

=== 3. BIDIRECTIONAL CONTRASTIVE PAIRWISE (FWD & REV) ===
SYSTEM:
You are reviewing two candidate RTL implementations (Candidate A and Candidate B) for functional conformance to the given specification.
Judge only the supplied RTL implementations. Do not write or repair RTL.
Exactly one implementation may conform cleanly, or both may fail/tie if evidence is ambiguous.
Return only valid JSON matching {"choice":"A|B|tie|neither","confidence":0.0,"reason":"short explanation"}.
Use one of the four literal choice strings ("A", "B", "tie", "neither"). Confidence must be between 0 and 1.

USER:
SPECIFICATION (verbatim from the existing dataset):
{specification}

CANDIDATE A:
```verilog
{candidate_a_rtl}
```

CANDIDATE B:
```verilog
{candidate_b_rtl}
```
Which candidate RTL is functionally consistent with the specification?
Output JSON only, with choice ("A", "B", "tie", or "neither"), confidence, and a short reason.

=== 4. BLINDED MULTI-JUDGE RATIONALE FIDELITY AUDIT ===
SYSTEM:
You are an impartial senior RTL verification engineer auditing an anonymous code reviewer's explanation on a KNOWN-BUGGY Verilog module.
Classify the reviewer's "reason" into EXACTLY ONE of:
- "grounded": Correctly identifies (partially or fully) the real functional defect or mismatched signal/timing behavior responsible for failing the spec/simulation. Minor wording imprecision is acceptable if the identified mechanism matches the real bug.
- "spurious_claim": Asserts a specific technical flaw in `buggy.sv` that is factually false, targets an unrelated/correct construct (e.g., falsely blaming non-blocking `<=`, nonexistent syntax/reset/latency error, or flipped diagnosis), while missing the actual defect.
- "vague": Contains only generic boilerplate ("violates specification", "incorrect logic", "syntax/timing errors") with no concrete defect localization.
Return ONLY valid JSON:
{"label":"grounded|spurious_claim|vague","actual_bug":"one sentence describing real bug","reviewer_claim_accurate":true,"confidence":0.0}

USER:
SPECIFICATION: {specification}
GOLDEN REFERENCE: {ref_rtl}
BUGGY IMPLEMENTATION: {buggy_rtl}
ICARUS VERILOG DIFF OUTPUT: {sim_log}
ANONYMOUS REVIEWER REASON TO AUDIT: {reviewer_reason}
\end{lstlisting}

\section{Benchmark Construction Details}
\label{app:construction}

\Cref{tab:funnel} gives the full curation funnel and \cref{tab:dataset_stats} the size statistics of each split.

\begin{table}[t]
\centering
\caption{Quantitative curation funnel across $448$ upstream rows ($363$ Core + $85$ Synthetic) yielding \textbf{Clean-114} ($228$ cases), \textbf{Artifact-6} ($12$ sensitivity cases), and \textbf{SynMut-50} ($100$ cases) across $126$ distinct designs.}
\label{tab:funnel}
\footnotesize
\setlength{\tabcolsep}{4pt}
\begin{tabularx}{\textwidth}{@{}l r >{\raggedright\arraybackslash}X@{}}
\toprule
\textbf{Pipeline Gate / Disposition} & \textbf{Count} & \textbf{Exclusion / Selection Basis (\cref{app:provenance})} \\
\midrule
\multicolumn{3}{l}{\textit{Core Track --- Source 1: ChipBench (MIT; Sequential Multi-Always Pool)}} \\
\quad Upstream debugging rows scanned & 89 & \code{state\_machine} (6), \code{timing} (29), \code{arithmetic} (24), \code{assignment} (30) \\
\quad \code{--} Excluded (incomplete spec / spec--ref conflict / non-seq) & $-36$ & Missing standalone specs ($-8$), conflicts ($-17$), comb.\ slice ($-11$) \\
\quad \code{--} Excluded by Icarus~11.0 compile / TB gate & $-4$ & Buggy compile fails ($-2$), golden reference TB mismatches ($-2$) \\
\quad \textbf{Eligible $\rightarrow$ Clean ChipBench Retained} & \textbf{49 $\rightarrow$ 17} & \textbf{17 seq / 0 comb} (\code{arithmetic}: 11, \code{assignment}: 6; $0$ artifacts) \\
\midrule
\multicolumn{3}{l}{\textit{Core Track --- Source 2: RTLRepairBench RealBugs (Apache-2.0 / MIT; VerilogEval~v2 Pool)}} \\
\quad Upstream RealBugs rows scanned & 274 & $112$ \textsc{VerilogEval}~v2 tasks ($252$ base + $22$ fine-tuned failures) \\
\quad \code{--} Leakage exclusions (spec hints / RTL comments) & $-8$ & Hint leakage in spec ($-6$), golden RTL correctness comments ($-2$) \\
\quad \code{--} Excluded by Icarus~11.0 compile / TB gate & $-15$ & Reference TB fails ($-9$), ref compile ($-5$), buggy SV cast ($-1$) \\
\quad Unique Core Designs Retained (\textbf{Full-120} Share) & $251 \rightarrow 103$ & \code{realbug\_base}: 93, \code{realbug\_tuned}: 10 \\
\quad \code{--} Audited Simulation-Only / \code{\#delay} Artifacts (\textbf{Artifact-6}) & $-6$ & $3$ time-$0$/\code{initial} (\code{P020}, \code{P047}, \code{P084}) + $3$ \code{\#delay} (\code{P018}, \code{P022}, \code{P028}) \\
\quad \textbf{Clean RTLRepairBench Retained} & \textbf{97} & \textbf{63 seq / 34 comb} (\code{realbug\_base}: 90, \code{realbug\_tuned}: 7) \\
\midrule
\multicolumn{3}{l}{\textit{Contrast Track --- SynMut-50 (RTLRepairBench Sem85 Synthetic Semantic Mutations)}} \\
\quad Upstream single-operator mutant rows scanned & 85 & \textsc{VerilogEval}~v2 \code{RefModule} + single-operator semantic mutations \\
\quad \code{--} Static leakage filter \& Icarus~11.0 differential gate & $85 \rightarrow 84$ & $1$ static check exclusion; all $84$ pairs ($168$ logs) verified by Icarus~11.0 \\
\quad \textbf{Balanced Rarest-Operator Deduplication} & \textbf{84 $\rightarrow$ 50} & \textbf{36 seq / 14 comb}; $6$ operators (\textbf{42} Clean RB overlap + $2$ Art.-6 + $6$ Syn-only) \\
\midrule
\textbf{Primary Clean / Sensitivity Corpus} & \textbf{114 + 50 (+6)} & \textbf{Clean-114 ($228$c) + SynMut-50 ($100$c)} [\textbf{Full-120 ($240$c) w/ Artifact-6}] \\
\bottomrule
\end{tabularx}
\end{table}

\begin{table}[t]
\centering
\caption{Descriptive structural statistics across \textbf{Clean-114}, excluded \textbf{Artifact-6}, \textbf{Full-120}, and \textbf{SynMut-50} over $126$ unique designs.}
\label{tab:dataset_stats}
\footnotesize
\setlength{\tabcolsep}{4pt}
\resizebox{\textwidth}{!}{%
\begin{tabular}{@{}lcccccccc@{}}
\toprule
\textbf{Dataset Slice} & \textbf{Pairs ($N$)} & \textbf{Cases ($2N$)} & \textbf{Seq / Comb} & \multicolumn{2}{c}{\textbf{RTL Lines of Code}} & \multicolumn{2}{c}{\textbf{Specification Words}} & \textbf{Bug-1st} \\
 & & & & \textbf{Mean} & \textbf{Med [Min--Max]} & \textbf{Mean} & \textbf{Med [Min--Max]} & \textbf{Order} \\
\midrule
\textsc{ChipBench} (Sequential, Clean) & 17 & 34 & 17 / 0 & 48.6 & 43.5 [11--107] & 249.9 & 254.0 [144--374] & 6 / 17 \\
\textsc{RTLRepairBench} Base (Clean) & 90 & 180 & 61 / 29 & 26.2 & 20.0 [6--134] & 148.2 & 117.5 [45--496] & 57 / 90 \\
\textsc{RTLRepairBench} Tuned (Clean) & 7 & 14 & 2 / 5 & 12.8 & 11.5 [6--27] & 85.7 & 65.0 [41--170] & 5 / 7 \\
Clean Sequential / Combinational & 80 / 34 & 160 / 68 & 80 / 34 & 33.5 / 17.5 & 27.0 / 13.0 & 175.2 / 122.6 & 144.5 / 110.0 & 47/80 ; 21/34 \\
\textbf{Clean-114 Primary Track} & \textbf{114} & \textbf{228} & \textbf{80 / 34} & \textbf{28.7} & \textbf{22.0 [6--134]} & \textbf{159.5} & \textbf{132.0 [41--496]} & \textbf{68 / 114} \\
\midrule
Excluded \textbf{Artifact-6} ($3$ sim + $3$ \code{\#delay}) & 6 & 12 & 2 / 4 & 9.8 & 8.5 [5--18] & 71.5 & 60.0 [38--145] & 4 / 6 \\
\textbf{Full-120 Sensitivity Core} & \textbf{120} & \textbf{240} & \textbf{82 / 38} & \textbf{27.8} & \textbf{21.0 [5--134]} & \textbf{155.1} & \textbf{124.5 [38--496]} & \textbf{72 / 120} \\
\midrule
\textbf{SynMut-50 Contrast Track} & \textbf{50} & \textbf{100} & \textbf{36 / 14} & \textbf{22.6} & \textbf{20.0 [9--58]} & \textbf{151.5} & \textbf{117.5 [42--496]} & \textbf{23 / 50} \\
\bottomrule
\end{tabular}%
}
\end{table}

\subsection{The Artifact-6 Oracle Audit}
\label{app:artifact6}

\paragraph{Oracle Artifact Audit (\textbf{Artifact-6} vs.\ \textbf{Clean-114}).} Post-run audit of low-mismatch cases, upstream \code{initial} statements, and explicit simulation delays identified \textbf{$6/120$ upstream pairs (\textbf{Artifact-6})} where testbench failure arises from simulator event semantics or inertial delays rather than synthesizable functional logic defects:
(a)~three time-$0$ / initialization artifacts where candidate RTL is synthesizably valid---\code{P020}/\code{C039} (\code{always @(*) out = in} omits the $t=0$ continuous-assignment event in Icarus, causing $2/100$ mismatches at $t=5$; all $12$ LLMs independently judge it \code{correct}), \code{P047}/\code{C094}, and \code{P084}/\code{C167} (upstream \code{RefModule}s use non-synthesizable \code{initial out=0}/\code{Q=0} before the first clock edge, $1/100$ and $1/199$ mismatches despite no reset in the spec); and
(b)~three \code{realbug\_tuned} inertial-delay cases (\code{P018}/\code{C035} \code{\#10}, \code{P022}/\code{C044} \code{\#10}, \code{P028}/\code{C055} \code{\#5}) whose Boolean logic is identical to \code{RefModule} but whose explicit Verilog \code{\#delay} lags the $5\,\text{ps}$ testbench clock half-period.
Across all $12$ reviewers, \textbf{Artifact-6} alone accounts for \textbf{$50.0\%$ ($59/118$) of all Full-120 False Acceptances} plus $4$ FPs. Rather than altering the frozen evaluation manifests, we retain \textbf{Full-120} as a sensitivity boundary and establish the filtered \textbf{Clean-114} corpus ($114$ pairs / $228$ cases: $80$ sequential, $34$ combinational; $17$ \textsc{ChipBench} + $90$ \code{realbug\_base} + $7$ \code{realbug\_tuned}; $68/114$ buggy-first) as our primary evaluation slice throughout \cref{sec:results}.

\subsection{Upstream Source Provenance and Exclusion Ledger}
\label{app:provenance}

\begin{table}[h]
\centering
\caption{Locked upstream repositories, commits, licenses, and curated \textbf{RTLReviewBench} pair counts across \textbf{Clean-114}, \textbf{Artifact-6}, and \textbf{SynMut-50}.}
\label{tab:provenance}
\footnotesize
\setlength{\tabcolsep}{4pt}
\begin{tabularx}{\textwidth}{@{}l >{\raggedright\arraybackslash}X l l r@{}}
\toprule
\textbf{Source Asset} & \textbf{Origin / Role} & \textbf{Commit SHA} & \textbf{License} & \textbf{Retained Pairs} \\
\midrule
\textsc{ChipBench}~\cite{yu2026chipbench} & Debug candidates + self-contained specs & \code{b4ba3a067598...} & MIT & 17 Clean (34c) \\
\textsc{RTLRepairBench} RealBugs~\cite{zhu2026rtlrepairbench} & LLM fails + \textsc{VerilogEval}~v2 refs & \code{17709198a893...} & Apache-2.0 / MIT & 97 Clean (+6 Art.) \\
\textsc{RTLRepairBench} Sem85~\cite{zhu2026rtlrepairbench} & Single-op mutants (\textbf{SynMut-50}) & \code{17709198a893...} & Apache-2.0 / MIT & 50 (100c) \\
\midrule
\textbf{Clean-114 + SynMut-50} & \textbf{$42$ Clean Matched ($+2$ Art.\ overlap, $6$ Syn-only)} & --- & --- & \textbf{164 (+6 Art.)} \\
\bottomrule
\end{tabularx}
\end{table}

\vspace{-3pt}
All candidate exclusions in \cref{tab:funnel,tab:provenance} execute deterministically:
\begin{itemize}[leftmargin=1.4em,itemsep=0.5pt,topsep=2pt]\small\raggedright
    \item \textbf{ChipBench ($89 \rightarrow 49 \rightarrow 17$):} $-8$ missing self-contained specs (\code{Prob018}, \code{Prob019}, \code{Prob021}, \code{Prob022}); $-17$ tasks exhibiting upstream spec--ref conflicts (\code{Prob005}, \code{Prob006}, \code{Prob007}, \code{Prob015}, \code{Prob023}, \code{Prob030}); $-11$ non-sequential tasks lacking \code{posedge clk} (\code{Prob011}, \code{Prob016}, \code{Prob017}, \code{Prob024}, \code{Prob025}); $-2$ buggy compile failures; $-2$ reference simulation mismatches. Unique-design deduplication yields $17$ Clean sequential pairs ($11$ \code{arithmetic}, $6$ \code{assignment}).
    \item \textbf{RTLRepairBench RealBugs ($274 \rightarrow 251 \rightarrow 103 \rightarrow 97$ Clean):} $-6$ specification hint leakage rows (\code{Prob062\_bugs\_mux2}, \code{Prob132\_always\_if2}); $-2$ reference comment leakage rows (\code{Prob114\_bugs\_case}); $-9$ reference testbench failures (\code{Prob034}, \code{Prob092}, \code{Prob094}); $-5$ reference compile port mismatches; $-1$ buggy SV cast error. Unique-design deduplication produces $103$ pairs, of which $6$ (\textbf{Artifact-6}: \code{P018}, \code{P020}, \code{P022}, \code{P028}, \code{P047}, \code{P084}) are documented as simulation-only/\code{\#delay} sensitivity cases, leaving $97$ Clean pairs ($90$ \code{realbug\_base}, $7$ \code{realbug\_tuned}; $63$ sequential, $34$ combinational).
    \item \textbf{RTLRepairBench Sem85 $\rightarrow$ \textbf{SynMut-50} ($85 \rightarrow 84 \rightarrow 50$):} $1$ static exclusion followed by Icarus~11.0 regression across $84$ pairs ($168$ logs). Balanced rarest-operator deduplication retains $50$ unique designs ($36$ seq, $14$ comb; $14$ \code{op\_swap}, $11$ \code{swap\_logical}, $10$ \code{swap\_ternary}, $10$ \code{flip\_equality}, $3$ \code{off\_by\_one\_literal}, $2$ \code{flip\_reset\_polarity}), sharing $42$ Clean overlapping designs (\textbf{Matched-42}) plus $2$ \textbf{Artifact-6} designs (\code{Prob010}, \code{Prob104}) with RealBugs.
\end{itemize}

\section{Full Results}
\label{app:full}

This appendix reports every metric behind the main-text summaries. \Cref{tab:models} lists the models and studies; \cref{tab:overall} gives the complete Clean-114 results with Full-120 sensitivity, calibration, and ensembles; \cref{tab:hard} the Hard split; \cref{tab:syn_vs_real} SynMut-50 and Matched-42; \cref{tab:interventions} Reason-First and pairwise review; \cref{tab:rf_ablation} the $2{\times}2$ ablation; \cref{tab:self_consistency} self-consistency; and \cref{tab:rationale_audit} the explanation audit.

\begin{table}[t]
\centering
\caption{Evaluated models ($12$ across $4$ paradigms) and complete empirical study matrix across \textbf{RTLReviewBench} ($17{,}623$ API queries total; $^\dagger$native/adaptive test-time \code{reasoning\_tokens}; counts shown as Clean-114 [Full-120]).}
\label{tab:models}
\footnotesize
\setlength{\tabcolsep}{4.0pt}
\resizebox{\textwidth}{!}{%
\begin{tabular}{@{}llcccccc@{}}
\toprule
\textbf{Model ID} & \textbf{Deployment Paradigm} & \textbf{Reasoning} & \textbf{Core} & \textbf{SynMut} & \textbf{Reason-} & \textbf{Pairwise /} & \textbf{SC@5} \\
 & & \textbf{Tokens$^\dagger$} & \textbf{228 [240]} & \textbf{(100)} & \textbf{First (228)} & \textbf{Audit (1,309)} & \textbf{(1,140)} \\
\midrule
\code{claude-haiku-5.5} & Budget Commercial (Anthropic) & None (0) & \checkmark & \checkmark & \checkmark & \checkmark / \checkmark & \checkmark \\
\code{gemini-3.8-flash}$^\dagger$ & Budget Commercial (Google) & Adaptive ($59\text{k}$--$766\text{k}$) & \checkmark & \checkmark & \checkmark & \checkmark / \checkmark & \checkmark \\
\code{gpt-4.1-mini} & Budget Commercial (OpenAI) & None (0) & \checkmark & \checkmark & \checkmark & \checkmark / \checkmark & --- \\
\midrule
\code{claude-4.6-sonnet} & Flagship Commercial (Anthropic) & Prose / Ext.\ (0) & \checkmark & \checkmark & \checkmark & \checkmark / \checkmark & --- \\
\code{gemini-3.1-pro}$^\dagger$ & Flagship Commercial (Google) & Native ($345\text{k}$ Clean) & \checkmark & \checkmark & \checkmark & \checkmark / \checkmark & --- \\
\code{gpt-5.4} & Flagship Commercial (OpenAI) & None (0) & \checkmark & \checkmark & \checkmark & \checkmark / \checkmark & \checkmark \\
\code{gpt-4.1} & Flagship Commercial (OpenAI) & None (0) & \checkmark & \checkmark & \checkmark & \checkmark / \checkmark & --- \\
\midrule
\code{claude-haiku-5.5-thinking}$^\dagger$ & Test-Time Reasoning (Anthropic) & Native ($156\text{k}$ Clean) & \checkmark & \checkmark & \checkmark & \checkmark / \checkmark & --- \\
\code{o4-mini}$^\dagger$ & Test-Time Reasoning (OpenAI) & Native ($325\text{k}$ Clean) & \checkmark & \checkmark & \checkmark & \checkmark / \checkmark & --- \\
\midrule
\code{gpt-oss-120b}$^\dagger$ & Open-Weight On-Premises & Native ($180\text{k}$ Clean) & \checkmark & \checkmark & \checkmark & \checkmark / \checkmark & --- \\
\code{gemma-4-31b-it} & Open-Weight On-Premises & None (0) & \checkmark & \checkmark & \checkmark & \checkmark / \checkmark & --- \\
\code{llama-3.3-70b-instruct} & Open-Weight On-Premises & None (0) & \checkmark & \checkmark & \checkmark & \checkmark / \checkmark & --- \\
\bottomrule
\end{tabular}%
}
\\[2pt]\raggedright\scriptsize{Note: Seven models report strictly $0$ API \code{reasoning\_tokens} across all extension studies; \code{gemini-3.8-flash} allocates adaptive hidden reasoning ($59.3\text{k}$ \code{SynMut-50}, $207.4\text{k}$ [$212.7\text{k}$ Full] \code{Reason-First}, $137.8\text{k}$ [$141.0\text{k}$] \code{Pairwise}, $765.9\text{k}$ [$786.4\text{k}$] \code{SC@5}), folded into \code{completion\_tokens} during Core (\cref{app:telemetry}). Study~3c ($2{\times}2$ \code{Reason-First} ablation: \code{vf\_trace} and \code{rf\_order\_only}) evaluates \code{gpt-4.1}, \code{gpt-4.1-mini}, and \code{gemma-4-31b-it} ($480$ queries each; $1{,}440$ total; $0$ \code{reasoning\_tokens}).}
\end{table}

\begin{table}[t]
\centering
\caption{Zero-shot functional conformance performance on the primary \textbf{Clean-114} track ($114$ designs / $228$ balanced cases) with exact $95\%$ Clopper-Pearson intervals~\cite{clopper1934use}, alongside \textbf{Full-120} ($240$c) sensitivity columns and cross-validated ensembles. \textbf{Bold} highlights top single-model and ensemble operating points.}
\label{tab:overall}
\footnotesize
\setlength{\tabcolsep}{2.45pt}
\resizebox{\textwidth}{!}{%
\begin{tabular}{@{}lccccccrr@{\hspace{4pt}}|@{\hspace{4pt}}cc@{}}
\toprule
\textbf{Model / Gate} & \textbf{Clean Acc.\ ($X/228$, 95\% CI)} & \textbf{FAR (Miss)} & \textbf{FRR (Over-Rej.)} & \textbf{Pair: Both / 1 / 0} & \textbf{Brier} & \textbf{Calib.} & \textbf{Med.\ Lat.} & \textbf{Med.\ Tok.} & \multicolumn{2}{c}{\textbf{Full-120 Sens.}} \\
 & & \textbf{($FN/114$)} & \textbf{($FP/114$)} & \textbf{($114$ Pairs)} & \textbf{Score} & \textbf{Gap} & \textbf{(ms)} & \textbf{(Ok / Err)} & \textbf{Acc.} & \textbf{FAR} \\
\midrule
\multicolumn{11}{l}{\textit{Single Models ($T=0.0$, Sorted by Clean-114 Strict Accuracy then Brier Score)}} \\
\code{claude-haiku-5.5-thk}$^\ddagger$ & \textbf{222/228 (97.4\%)} {\scriptsize[94.4, 99.0]} & 3/114 (2.6\%) & \textbf{2/114 (1.8\%$^\ddagger$)} & \textbf{108} / 6 / 0 (94.7\%) & \textbf{0.0280} & $-7.4\pp$ & 8998 & 555.0 / 1403.5 & 95.0\% & 7.5\% \\
\code{claude-haiku-5.5} & \textbf{222/228 (97.4\%)} {\scriptsize[94.4, 99.0]} & 4/114 (3.5\%) & \textbf{2/114 (1.8\%)} & \textbf{108} / 6 / 0 (94.7\%) & 0.0310 & $-7.2\pp$ & 3617 & 476.5 / 1595.0 & 95.0\% & 8.3\% \\
\code{gemini-3.1-pro} & 221/228 (96.9\%) {\scriptsize[93.8, 98.8]} & 1/114 (0.9\%) & 6/114 (5.3\%) & 107 / 7 / 0 (93.9\%) & 0.0299 & $+3.0\pp$ & 12237 & 903.0 / 1878.0 & 95.0\% & 5.0\% \\
\code{gemini-3.8-flash} & 213/228 (93.4\%) {\scriptsize[89.4, 96.3]} & \textbf{0/114 (0.0\%)} & 15/114 (13.2\%) & 99 / 15 / 0 (86.8\%) & 0.0590 & $+5.3\pp$ & 5505 & 431.0 / 1218.0 & 92.5\% & 2.5\% \\
\code{claude-4.6-sonnet}$^\S$ & 212/228 (93.0\%) {\scriptsize[88.9, 95.9]} & 3/114 (2.6\%) & 13/114 (11.4\%) & 100 / 12 / 2 (87.7\%) & 0.0600 & \textbf{$+2.1\pp$} & 8530 & 383.5 / 625.5 & 91.2\% & 6.7\% \\
\code{gpt-oss-120b} & 209/228 (91.7\%) {\scriptsize[87.3, 94.9]} & 2/114 (1.8\%) & 17/114 (14.9\%) & 96 / 17 / 1 (84.2\%) & 0.0727 & $+2.8\pp$ & 8815 & 620.0 / 953.0 & 90.0\% & 5.8\% \\
\code{o4-mini} & 207/228 (90.8\%) {\scriptsize[86.3, 94.2]} & 3/114 (2.6\%) & 18/114 (15.8\%) & 94 / 19 / 1 (82.5\%) & 0.1048 & $-8.8\pp$ & 8959 & 874.0 / 2302.0 & 89.2\% & 6.7\% \\
\code{gpt-5.4} & 191/228 (83.8\%) {\scriptsize[78.3, 88.3]} & 4/114 (3.5\%) & 33/114 (28.9\%) & 79 / 33 / 2 (69.3\%) & 0.1558 & $+14.5\pp$ & 4076 & 70.0 / 82.0 & 82.1\% & 7.5\% \\
\code{gemma-4-31b-it} & 180/228 (78.9\%) {\scriptsize[73.1, 84.1]} & 4/114 (3.5\%) & 44/114 (38.6\%) & 66 / 48 / 0 (57.9\%) & 0.2100 & $+20.8\pp$ & 3029 & 84.0 / 212.0 & 77.5\% & 7.5\% \\
\code{gpt-4.1-mini} & 164/228 (71.9\%) {\scriptsize[65.6, 77.7]} & 3/114 (2.6\%) & 61/114 (53.5\%) & 52 / 60 / 2 (45.6\%) & 0.2348 & $+19.6\pp$ & \textbf{2158} & 92.5 / 126.0 & 71.2\% & 5.8\% \\
\code{gpt-4.1} & 159/228 (69.7\%) {\scriptsize[63.3, 75.6]} & 2/114 (1.8\%) & 67/114 (58.8\%) & 46 / 67 / 1 (40.4\%) & 0.2716 & $+25.5\pp$ & 6247 & 79.0 / 99.0 & 68.8\% & 5.8\% \\
\code{llama-3.3-70b-instruct} & 151/228 (66.2\%) {\scriptsize[59.7, 72.3]} & 30/114 (26.3\%) & 47/114 (41.2\%) & 41 / 69 / 4 (36.0\%) & 0.3278 & $+32.3\pp$ & 3707 & \textbf{29.0 / 27.0} & 65.8\% & 29.2\% \\
\midrule
\multicolumn{11}{l}{\textit{Multi-Model Boolean Gates, 5-Fold Cross-Validation, \& Deferral Regimes}} \\
Top-3 Diverse Maj.\ (H, P31, Oss) & \textbf{224/228 (98.2\%)} {\scriptsize[95.6, 99.5]} & 1/114 (0.9\%) & 3/114 (2.6\%) & \textbf{110} / 4 / 0 (96.5\%) & --- & --- & \multicolumn{1}{c}{---} & \multicolumn{1}{c@{\hspace{4pt}}|@{\hspace{4pt}}}{---} & 96.2\% & 5.0\% \\
5-Fold Pair CV (Top-3 Fam.\ / Any) & 224 (98.2\%) / 222 (97.4\%) & 1 (0.9\%) / 1 (0.9\%) & 3 (2.6\%) / 5 (4.4\%) & --- & --- & --- & \multicolumn{1}{c}{---} & \multicolumn{1}{c@{\hspace{4pt}}|@{\hspace{4pt}}}{---} & \multicolumn{1}{c}{---} & \multicolumn{1}{c}{---} \\
Top-3 Comm.\ Maj.\ (H, P31, Son) & 221/228 (96.9\%) {\scriptsize[93.8, 98.8]} & 2/114 (1.8\%) & 5/114 (4.4\%) & 107 / 7 / 0 (93.9\%) & --- & --- & \multicolumn{1}{c}{---} & \multicolumn{1}{c@{\hspace{4pt}}|@{\hspace{4pt}}}{---} & 95.0\% & 5.8\% \\
Low-FP AND Gate (H \& P31) & 223/228 (97.8\%) {\scriptsize[94.9, 99.3]} & 4/114 (3.5\%) & \textbf{1/114 (0.9\%)} & 109 / 5 / 0 (95.6\%) & --- & --- & \multicolumn{1}{c}{---} & \multicolumn{1}{c@{\hspace{4pt}}|@{\hspace{4pt}}}{---} & 95.4\% & 8.3\% \\
Strict OR Gate (H $\vert$ Fl) & 212/228 (93.0\%) {\scriptsize[88.9, 95.9]} & \textbf{0/114 (0.0\%)} & 16/114 (14.0\%) & 98 / 16 / 0 (86.0\%) & --- & --- & \multicolumn{1}{c}{---} & \multicolumn{1}{c@{\hspace{4pt}}|@{\hspace{4pt}}}{---} & 92.1\% & 2.5\% \\
Consensus Subset (H == P31)$^\P$ & 217/219 (99.1\%) {\scriptsize[96.7, 99.9]} & 1/111 (0.9\%) & 1/108 (0.9\%) & --- ($96.1\%$ cov.) & --- & --- & \multicolumn{1}{c}{---} & \multicolumn{1}{c@{\hspace{4pt}}|@{\hspace{4pt}}}{---} & 97.0\% & 5.2\% \\
Consensus Subset (H == Fl)$^\P$ & \textbf{208/209 (99.5\%)} {\scriptsize[97.4, 100.0]} & \textbf{0/110 (0.0\%)} & 1/99 (1.0\%) & --- ($91.7\%$ cov.) & --- & --- & \multicolumn{1}{c}{---} & \multicolumn{1}{c@{\hspace{4pt}}|@{\hspace{4pt}}}{---} & 98.2\% & 2.7\% \\
\bottomrule
\end{tabular}%
}
\\[2pt]\raggedright\scriptsize{$^\ddagger$$99.6\%$ coverage ($227/228$; valid case \code{C206} failed JSON quotes $\rightarrow 2$ \code{FP} + $1$ \code{I-1} = $3/114$ valid-case errors; selective acc $97.8\%$ [$94.9, 99.3$]). $^\S$$33.3\%$ Clean ($36.2\%$ Full) strict JSON / $100\%$ recovery ($11/13$ FPs in prose-prefixed CoT vs.\ $2/76$ in strict JSON). $^\P$Accuracy on agreement slice; remaining disagreements ($9$ for H==P31, $19$ for H==Fl) routed to EDA simulation. Abbreviations: H=\code{claude-haiku-5.5}, P31=\code{gemini-3.1-pro}, Fl=\code{gemini-3.8-flash}, Oss=\code{gpt-oss-120b}, Son=\code{claude-4.6-sonnet}.}
\end{table}

\begin{table}[t]
\centering
\caption{\textbf{RTLReviewBench-Hard} ($22$ Clean pairs / $44$ cases; selection: $\ge 6/12$ reviewers misclassify a sibling under \code{Verdict-First}). Scores for these $12$ models are in-sample by construction.}
\label{tab:hard}
\footnotesize
\setlength{\tabcolsep}{4pt}
\begin{tabular}{@{}lcccccc@{}}
\toprule
\textbf{Model} & \textbf{Clean-114 Acc.} & \textbf{Acc.} & \textbf{FAR} & \textbf{FRR} & \textbf{Pair} & \textbf{Reason-First Acc.} \\
\midrule
\code{claude-haiku-5.5} & 97.4\% & \textbf{88.6\%} & 18.2\% & \textbf{4.5\%} & \textbf{77.3\%} & \textbf{90.9\%} \\
\code{claude-haiku-5.5-thinking} & 97.4\% & \textbf{88.6\%} & 13.6\% & 9.1\% & \textbf{77.3\%} & \textbf{90.9\%} \\
\code{gemini-3.1-pro} & 96.9\% & 84.1\% & 4.5\% & 27.3\% & 68.2\% & \textbf{90.9\%} \\
\code{gpt-oss-120b} & 91.7\% & 75.0\% & 9.1\% & 40.9\% & 54.5\% & 72.7\% \\
\code{gemini-3.8-flash} & 93.4\% & 72.7\% & \textbf{0.0\%} & 54.5\% & 45.5\% & 77.3\% \\
\code{claude-4.6-sonnet} & 93.0\% & 70.5\% & 13.6\% & 45.5\% & 50.0\% & 77.3\% \\
\code{o4-mini} & 90.8\% & 70.5\% & 13.6\% & 45.5\% & 45.5\% & 81.8\% \\
\code{gemma-4-31b-it} & 78.9\% & 52.3\% & 13.6\% & 81.8\% & 4.5\% & 63.6\% \\
\code{gpt-4.1} & 69.7\% & 50.0\% & 9.1\% & 90.9\% & 4.5\% & 65.9\% \\
\code{gpt-5.4} & 83.8\% & 50.0\% & 13.6\% & 86.4\% & 9.1\% & 56.8\% \\
\code{llama-3.3-70b-instruct} & 66.2\% & 50.0\% & 22.7\% & 77.3\% & 9.1\% & 52.3\% \\
\code{gpt-4.1-mini} & 71.9\% & 47.7\% & 9.1\% & 95.5\% & 4.5\% & 63.6\% \\
\bottomrule
\end{tabular}
\end{table}

\begin{table}[t]
\centering
\caption{\textbf{SynMut-50} ($100$ cases / $50$ designs) vs.\ Clean \textbf{Matched-42} real-vs-synthetic performance across all $12$ models ($95\%$ Clopper-Pearson CIs), plus pooled SynMut-50 miss breakdown across six operators ($600$ buggy evaluations).}
\label{tab:syn_vs_real}
\footnotesize
\setlength{\tabcolsep}{2.9pt}
\resizebox{\textwidth}{!}{%
\begin{tabular}{@{}lccccccc@{}}
\toprule
\textbf{Model} & \multicolumn{3}{c}{\textbf{SynMut-50 Full Track ($N=100$)}} & \multicolumn{4}{c}{\textbf{Clean Matched-42 Base Designs ($84$ Cases Each)}} \\
\cmidrule(lr){2-4}\cmidrule(lr){5-8}
 & \textbf{Acc.\ (95\% CI)} & \textbf{FAR ($/50$)} & \textbf{FRR ($/50$)} & \textbf{Real $\rightarrow$ Syn Acc.} & \textbf{$p_{\text{raw}}$ / $p_{\text{Holm}}$} & \textbf{Real $\rightarrow$ Syn FAR} & \textbf{Real $\rightarrow$ Syn FRR} \\
\midrule
\code{claude-haiku-5.5-thinking} & \textbf{99.0\%} {\scriptsize[94.6,100]} & \textbf{0.0\% (0)} & \textbf{2.0\% (1)} & $95.2\% \rightarrow 98.8\%$ & $0.250$ / $1.00$ & $2.4\% \rightarrow 0.0\%$ & $7.1\% \rightarrow 2.4\%$ \\
\code{gemini-3.1-pro} & \textbf{99.0\%} {\scriptsize[94.6,100]} & \textbf{0.0\% (0)} & \textbf{2.0\% (1)} & $96.4\% \rightarrow 98.8\%$ & $0.500$ / $1.00$ & $2.4\% \rightarrow 0.0\%$ & $4.8\% \rightarrow 2.4\%$ \\
\code{claude-haiku-5.5} & 97.0\% {\scriptsize[91.5, 99.4]} & 2.0\% (1) & 4.0\% (2) & $96.4\% \rightarrow 97.6\%$ & $1.000$ / $1.00$ & $4.8\% \rightarrow 0.0\%$ & $2.4\% \rightarrow 4.8\%$ \\
\code{gemini-3.8-flash} & 96.0\% {\scriptsize[90.1, 98.9]} & 2.0\% (1) & 6.0\% (3) & $95.2\% \rightarrow 95.2\%$ & $1.000$ / $1.00$ & $0.0\% \rightarrow 2.4\%$ & $9.5\% \rightarrow 7.1\%$ \\
\code{o4-mini} & 96.0\% {\scriptsize[90.1, 98.9]} & \textbf{0.0\% (0)} & 8.0\% (4) & $89.3\% \rightarrow 96.4\%$ & $0.031$ / $0.375$ & $7.1\% \rightarrow 0.0\%$ & $14.3\% \rightarrow 7.1\%$ \\
\code{claude-4.6-sonnet} & 94.0\% {\scriptsize[87.4, 97.8]} & 2.0\% (1) & 10.0\% (5) & $91.7\% \rightarrow 94.0\%$ & $0.625$ / $1.00$ & $4.8\% \rightarrow 2.4\%$ & $11.9\% \rightarrow 9.5\%$ \\
\code{gpt-oss-120b} & 87.0\% {\scriptsize[78.8, 92.9]} & \textbf{0.0\% (0)} & 26.0\% (13) & $88.1\% \rightarrow 88.1\%$ & $1.000$ / $1.00$ & $4.8\% \rightarrow 0.0\%$ & $19.0\% \rightarrow 23.8\%$ \\
\code{gpt-5.4} & 85.0\% {\scriptsize[76.5, 91.4]} & 2.0\% (1) & 28.0\% (14) & $82.1\% \rightarrow 85.7\%$ & $0.375$ / $1.00$ & $7.1\% \rightarrow 2.4\%$ & $28.6\% \rightarrow 26.2\%$ \\
\code{gemma-4-31b-it} & 82.0\% {\scriptsize[73.1, 89.0]} & \textbf{0.0\% (0)} & 36.0\% (18) & $78.6\% \rightarrow 83.3\%$ & $0.125$ / $1.00$ & $4.8\% \rightarrow 0.0\%$ & $38.1\% \rightarrow 33.3\%$ \\
\code{llama-3.3-70b-instruct} & 81.0\% {\scriptsize[71.9, 88.2]} & 8.0\% (4) & 30.0\% (15) & $73.8\% \rightarrow 81.0\%$ & $0.109$ / $1.00$ & $21.4\% \rightarrow 9.5\%$ & $31.0\% \rightarrow 28.6\%$ \\
\code{gpt-4.1} & 76.0\% {\scriptsize[66.4, 84.0]} & 2.0\% (1) & 46.0\% (23) & $77.4\% \rightarrow 75.0\%$ & $0.625$ / $1.00$ & $2.4\% \rightarrow 2.4\%$ & $42.9\% \rightarrow 47.6\%$ \\
\code{gpt-4.1-mini} & 73.0\% {\scriptsize[63.2, 81.4]} & 2.0\% (1) & 52.0\% (26) & $70.2\% \rightarrow 71.4\%$ & $1.000$ / $1.00$ & $4.8\% \rightarrow 2.4\%$ & $54.8\% \rightarrow 54.8\%$ \\
\midrule
\multicolumn{4}{l}{\textit{Pooled Across All 12 Models (Matched-42: $504$ Buggy / $504$ Valid)}} & $86.2\% \rightarrow 88.8\%$ & --- & \textbf{5.6\% $\rightarrow$ 1.8\%} & $22.0\% \rightarrow 20.6\%$ \\
\bottomrule
\end{tabular}%
}
\\[3pt]\scriptsize{\textbf{Pooled SynMut-50 Miss Breakdown ($10/600 = 1.7\%$ [$0.8, 3.0$]):} \code{swap\_ternary} $6/120$ ($5.0\%$ [$1.9, 10.6$]); \code{flip\_reset\_polarity} $1/24$ ($4.2\%$ [$0.1, 21.1$]); \code{off\_by\_one\_literal} $1/36$ ($2.8\%$ [$0.1, 14.5$]); \code{flip\_equality} $1/120$ ($0.8\%$ [$0.0, 4.6$]); \code{op\_swap} $1/168$ ($0.6\%$ [$0.0, 3.3$]); \code{swap\_logical} $0/132$ ($0.0\%$ [$0.0, 2.8$]). Matched-42 pooled \textsc{FAR}: $28/504$ [$3.7, 7.9$] $\rightarrow 9/504$ [$0.8, 3.4$].}
\end{table}

\begin{table}[t]
\centering
\caption{Complete 12-model \code{Reason-First} vs.\ \code{Verdict-First} prompting comparison ($N=228$ Clean cases each) and Bidirectional Contrastive \code{Pairwise} results ($114$ Clean pairs $\times 2$ directions) with Holm--Bonferroni adjusted $p$-values.}
\label{tab:interventions}
\footnotesize
\setlength{\tabcolsep}{2.7pt}
\resizebox{\textwidth}{!}{%
\begin{tabular}{@{}lccccccc|cccccc@{}}
\toprule
\textbf{Model} & \multicolumn{7}{c|}{\textbf{Prompt Intervention: Verdict-First $\rightarrow$ Reason-First (Clean $N=228$)}} & \multicolumn{6}{c}{\textbf{Contrastive Pairwise (Clean $114$ Pairs $\times 2$)}} \\
\cmidrule(lr){2-8}\cmidrule(lr){9-14}
 & \textbf{Acc.\ (VF$\rightarrow$RF)} & \textbf{$\Delta$Acc} & \textbf{$p_{\text{raw}}$ / $p_{\text{Holm}}$} & \textbf{FAR (VF$\rightarrow$RF)} & \textbf{FRR (VF$\rightarrow$RF)} & \textbf{Brier} & \textbf{Tok.} & \textbf{Pool} & \textbf{Swap} & \textbf{$\Delta$Iso} & \textbf{Flips (A/B)} & \textbf{Bias} & \textbf{Neith.} \\
\midrule
\code{gpt-4.1} & $69.7 \rightarrow \mathbf{86.0\%}$ & \textbf{$+16.2\pp$} & $4.3\text{e-}7$ / $\mathbf{5.2\text{e-}6}$ & $1.8 \rightarrow 8.8\%$ & $58.8 \rightarrow \mathbf{19.3\%}$ & $0.272 \rightarrow 0.135$ & $79 \rightarrow 260$ & $72.8\%$ & $62.3\%$ & \textbf{$+21.9\pp^{\ast\ast\ast}$} & \textbf{6 (5/1)} & \textbf{$+10.5\pp$} & 40 \\
\code{gpt-4.1-mini} & $71.9 \rightarrow \mathbf{83.8\%}$ & \textbf{$+11.8\pp$} & $5.8\text{e-}4$ / $\mathbf{0.0058}$ & $2.6 \rightarrow 14.0\%$ & $53.5 \rightarrow \mathbf{18.4\%}$ & $0.235 \rightarrow 0.148$ & $93 \rightarrow 296$ & $66.7\%$ & $52.6\%$ & $+7.0\pp$ & \textbf{11 (9/2)} & $+7.5\pp$ & \textbf{46} \\
\code{gemma-4-31b-it} & $78.9 \rightarrow \mathbf{90.4\%}$ & \textbf{$+11.4\pp$} & $1.3\text{e-}5$ / $\mathbf{1.4\text{e-}4}$ & $\mathbf{3.5 \rightarrow 3.5\%}$ & $38.6 \rightarrow \mathbf{15.8\%}$ & $0.210 \rightarrow 0.094$ & $84 \rightarrow 300$ & $82.9\%$ & $76.3\%$ & \textbf{$+18.4\pp^{\ast\ast\ast}$} & $3$ ($3$/$0$) & $+1.8\pp$ & 34 \\
\code{gpt-5.4} & $83.8 \rightarrow 88.6\%$ & $+4.8\pp$ & $0.043$ / $0.390$ & $3.5 \rightarrow 3.5\%$ & $28.9 \rightarrow 19.3\%$ & $0.156 \rightarrow 0.107$ & $70 \rightarrow 320$ & $78.5\%$ & $72.8\%$ & $+3.5\pp$ & $1$ ($1$/$0$) & $-0.9\pp$ & \textbf{43} \\
\code{o4-mini} & $90.8 \rightarrow 93.9\%$ & $+3.1\pp$ & $0.065$ / $0.523$ & $2.6 \rightarrow 1.8\%$ & $15.8 \rightarrow 10.5\%$ & $0.105 \rightarrow 0.096$ & $874 \rightarrow 1130$ & $92.1\%$ & $89.5\%$ & $+7.0\pp$ & $1$ ($1$/$0$) & $+2.2\pp$ & 14 \\
\code{gemini-3.1-pro} & $96.9 \rightarrow \mathbf{98.2\%}$ & $+1.3\pp$ & $0.250$ / $1.00$ & $\mathbf{0.9 \rightarrow 0.0\%}$ & $5.3 \rightarrow 3.5\%$ & $\mathbf{0.030 \rightarrow 0.018}$ & $903 \rightarrow 1204$ & \textbf{98.7\%} & \textbf{98.2\%} & $+4.4\pp$ & $0$ ($0$/$0$) & $-0.4\pp$ & 3 \\
\code{llama-3.3-70b} & $66.2 \rightarrow 67.1\%$ & $+0.9\pp$ & $0.864$ / $1.00$ & $26.3 \rightarrow 36.0\%$ & $41.2 \rightarrow 28.9\%$ & $0.328 \rightarrow 0.290$ & $29 \rightarrow 255$ & $64.9\%$ & $52.6\%$ & \textbf{$+16.7\pp^\ast$} & \textbf{20 (16/4)} & \textbf{$+11.4\pp$} & 16 \\
\code{claude-4.6-sonnet}$^\ddagger$ & $93.0 \rightarrow 93.4\%$ & $+0.4\pp$ & $1.00$ / $1.00$ & $2.6 \rightarrow 2.6\%$ & $11.4 \rightarrow 10.5\%$ & $0.060 \rightarrow 0.056$ & $384 \rightarrow 588$ & $96.5\%$ & $95.6\%$ & $+7.9\pp$ & $0$ ($0$/$0$) & $-0.9\pp$ & 6 \\
\code{gemini-3.8-flash}$^\mathsection$ & $93.4 \rightarrow 93.9\%$ & $+0.4\pp$ & $1.00$ / $1.00$ & $\mathbf{0.0 \rightarrow 0.0\%}$ & $13.2 \rightarrow 12.3\%$ & $0.059 \rightarrow 0.056$ & $431 \rightarrow 679$ & \textbf{98.7\%} & $97.4\%$ & \textbf{$+10.5\pp^{\ast\ast}$} & $0$ ($0$/$0$) & $+0.4\pp$ & 3 \\
\code{claude-haiku-5.5} & $97.4 \rightarrow 97.4\%$ & $0.0\pp$ & $1.00$ / $1.00$ & $3.5 \rightarrow 1.8\%$ & $1.8 \rightarrow 3.5\%$ & $0.031 \rightarrow 0.032$ & $477 \rightarrow 688$ & $95.2\%$ & $93.9\%$ & $-0.9\pp$ & $1$ ($0$/$1$) & $-0.9\pp$ & 5 \\
\code{claude-haiku-5.5-thk}$^\dagger$ & $97.4 \rightarrow 97.4\%$ & $0.0\pp$ & $1.00$ / $1.00$ & $2.6 \rightarrow 0.9\%$ & $\mathbf{1.8 \rightarrow 2.6\%}$ & $0.028 \rightarrow 0.028$ & $555 \rightarrow 765$ & $96.5\%$ & $94.7\%$ & $+0.0\pp$ & $0$ ($0$/$0$) & $+0.9\pp$ & 4 \\
\code{gpt-oss-120b} & $91.7 \rightarrow 90.4\%$ & $-1.3\pp$ & $0.549$ / $1.00$ & $1.8 \rightarrow 1.8\%$ & $14.9 \rightarrow 17.5\%$ & $0.073 \rightarrow 0.081$ & $620 \rightarrow 756$ & $85.1\%$ & $78.1\%$ & $-6.1\pp$ & $0$ ($0$/$0$) & $+1.3\pp$ & 29 \\
\bottomrule
\end{tabular}%
}
\\[2pt]\raggedright\scriptsize{Note: Clean \code{Reason-First} FRR drops cleanly under Holm correction for \code{gpt-4.1} ($p_{\text{Holm}}=8.2\times 10^{-12}$), \code{gpt-4.1-mini} ($p_{\text{Holm}}=1.2\times 10^{-9}$), \code{gemma-4} ($p_{\text{Holm}}=6.2\times 10^{-5}$), and \code{llama-3.3} ($p_{\text{Holm}}=0.040$, while \code{gpt-4.1-mini} FAR rises $p_{\text{Holm}}=0.012$); \code{llama-3.3} emits $1$ valid abstain (\code{C187}). $^\dagger$\code{haiku-5.5-thinking} had $2$ invalids under RF ($99.1\%$ cov.) and Pairwise. $^\ddagger$\code{sonnet} had $40.4\%$ Clean RF / $17.5\%$ PW strict JSON. $^\mathsection$\code{gem-3.8-flash} logged $207.4\text{k}$/$137.8\text{k}$ Clean adaptive reasoning tokens on RF/PW. On \code{Pairwise}, $^\ast$ denotes Holm-adjusted significance vs.\ Isolated Joint-Pair Acc ($^\ast p_{\text{Holm}}<0.05$, $^{\ast\ast}p_{\text{Holm}}<0.01$, $^{\ast\ast\ast}p_{\text{Holm}}<0.001$): \code{gemma-4} $p_{\text{Holm}}=2.3\times 10^{-4}$, \code{gpt-4.1} $p_{\text{Holm}}=4.5\times 10^{-4}$, \code{gem-3.8-flash} $p_{\text{Holm}}=0.0049$, \code{llama-3.3} $p_{\text{Holm}}=0.012$ (\code{sonnet} $p_{\text{raw}}=0.022/p_{\text{Holm}}=0.18$, \code{o4-mini} $p_{\text{raw}}=0.039/p_{\text{Holm}}=0.27$, \code{gem-3.1-pro} $p_{\text{raw}}=0.063/p_{\text{Holm}}=0.38$).}
\end{table}

\begin{table}[t]
\centering
\caption{Study~3c $2{\times}2$ factorial \code{Reason-First} ablation (\code{vf\_trace} and \code{rf\_order\_only}; $1{,}440$ new greedy $T=0.0$ queries) separating JSON schema key order from step-by-step tracing on the three largest \code{Reason-First} beneficiaries (\textbf{Clean-114} primary, $N=228$ cases/cell).}
\label{tab:rf_ablation}
\footnotesize
\setlength{\tabcolsep}{3.0pt}
\resizebox{\textwidth}{!}{%
\begin{tabular}{@{}llccccccc@{}}
\toprule
\textbf{Model} & \textbf{Cell (Order, Trace)} & \textbf{Clean Acc.\ (95\% CI)} & \textbf{FAR} & \textbf{FRR} & \textbf{Pair-Both} & \textbf{Brier} & \textbf{Tok.$^\S$} & \textbf{$\Delta$VF ($p_{\text{Holm}}$)} \\
\midrule
\code{gpt-4.1} & \code{VF} (\code{V$\rightarrow$C$\rightarrow$R}, no) & 159/228 (69.7\%) {\scriptsize[63.3, 75.6]} & \textbf{1.8\%} & 58.8\% & 40.4\% & 0.2716 & 83.0 & --- \\
 & \code{A: vf\_trace} (\code{V$\rightarrow$C$\rightarrow$R}, yes) & 153/228 (67.1\%) {\scriptsize[60.6, 73.2]} & 3.5\% & 61.4\% & 36.8\% & 0.2994 & 238.0 & $-2.6\pp$ ($0.438$) \\
 & \code{B: rf\_order\_only} (\code{R$\rightarrow$V$\rightarrow$C}, no) & 177/228 (77.6\%) {\scriptsize[71.7, 82.9]} & 10.5\% & 33.3\% & 58.8\% & 0.2017 & 110.0 & \textbf{$+7.9\pp$ ($0.048^\ast$)} \\
 & \code{RF} (\code{R$\rightarrow$V$\rightarrow$C}, yes) & \textbf{196/228 (86.0\%)} {\scriptsize[80.8, 90.2]} & 8.8\% & \textbf{19.3\%} & \textbf{73.7\%} & \textbf{0.1346} & 270.5 & $+16.2\pp$ ($5.2\text{e-}6^\dagger$) \\
\midrule
\code{gpt-4.1-mini} & \code{VF} (\code{V$\rightarrow$C$\rightarrow$R}, no) & 164/228 (71.9\%) {\scriptsize[65.6, 77.7]} & \textbf{2.6\%} & 53.5\% & 45.6\% & 0.2348 & 100.0 & --- \\
 & \code{A: vf\_trace} (\code{V$\rightarrow$C$\rightarrow$R}, yes) & 157/228 (68.9\%) {\scriptsize[62.4, 74.8]} & 3.5\% & 58.8\% & 40.4\% & 0.2618 & 315.0 & $-3.1\pp$ ($0.438$) \\
 & \code{B: rf\_order\_only} (\code{R$\rightarrow$V$\rightarrow$C}, no) & 186/228 (81.6\%) {\scriptsize[75.9, 86.4]} & 13.2\% & 23.7\% & 67.5\% & 0.1611 & 105.0 & \textbf{$+9.6\pp$ ($0.017^\ast$)} \\
 & \code{RF} (\code{R$\rightarrow$V$\rightarrow$C}, yes) & \textbf{191/228 (83.8\%)} {\scriptsize[78.3, 88.3]} & 14.0\% & \textbf{18.4\%} & \textbf{70.2\%} & \textbf{0.1477} & 302.0 & $+11.8\pp$ ($0.0058^\dagger$) \\
\midrule
\code{gemma-4-31b-it} & \code{VF} (\code{V$\rightarrow$C$\rightarrow$R}, no) & 180/228 (78.9\%) {\scriptsize[73.1, 84.1]} & 3.5\% & 38.6\% & 57.9\% & 0.2100 & 95.0 & --- \\
 & \code{A: vf\_trace}$^\ddagger$ (\code{V$\rightarrow$C$\rightarrow$R}, yes) & 184/228 (80.7\%) {\scriptsize[75.0, 85.6]} & \textbf{1.8\%} & 34.2\% & 62.3\% & 0.1823 & 304.5 & $+1.8\pp$ ($0.688$) \\
 & \code{B: rf\_order\_only} (\code{R$\rightarrow$V$\rightarrow$C}, no) & 194/228 (85.1\%) {\scriptsize[79.8, 89.4]} & 2.6\% & 27.2\% & 71.9\% & 0.1444 & 138.5 & \textbf{$+6.1\pp$ ($0.046^\ast$)} \\
 & \code{RF} (\code{R$\rightarrow$V$\rightarrow$C}, yes) & \textbf{206/228 (90.4\%)} {\scriptsize[85.8, 93.9]} & 3.5\% & \textbf{15.8\%} & \textbf{82.5\%} & \textbf{0.0940} & 315.0 & $+11.4\pp$ ($1.4\text{e-}4^\dagger$) \\
\midrule
\multicolumn{2}{@{}l}{\textit{Accuracy Decomposition ($\pp$)}} & \textbf{Total ($\mathrm{RF}-\mathrm{VF}$)} & \textbf{Order ($\mathrm{B}-\mathrm{VF}$)} & \textbf{Instr.\ ($\mathrm{A}-\mathrm{VF}$)} & \textbf{Inter.} & \textbf{$p_{\text{H}}$(\code{RF} vs.\ \code{A})} & \textbf{$p_{\text{H}}$(\code{RF} vs.\ \code{B})} & \textbf{Full-120 (O / I / Int)} \\
\midrule
\multicolumn{2}{@{}l}{\code{gpt-4.1}} & $+16.2\pp$ & $+7.9\pp$ & $-2.6\pp$ & $+11.0\pp$ & $5.1\text{e-}8^{\ast\ast\ast}$ & $0.012^\ast$ & $+16.2$ ($+7.9/{-}1.7/+10.0$) \\
\multicolumn{2}{@{}l}{\code{gpt-4.1-mini}} & $+11.8\pp$ & $+9.6\pp$ & $-3.1\pp$ & $+5.3\pp$ & $1.9\text{e-}4^{\ast\ast\ast}$ & $0.688$ & $+11.2$ ($+9.2/{-}2.9/+5.0$) \\
\multicolumn{2}{@{}l}{\code{gemma-4-31b-it}} & $+11.4\pp$ & $+6.1\pp$ & $+1.8\pp$ & $+3.5\pp$ & $0.0031^{\ast\ast}$ & $0.046^\ast$ & $+11.2$ ($+6.2/{+}2.1/+2.9$) \\
\multicolumn{2}{@{}l}{\textbf{Mean (3 Models)}} & \textbf{$+13.2\pp$} & \textbf{$+7.9\pp$} & \textbf{$-1.3\pp$} & \textbf{$+6.6\pp$} & --- & --- & \textbf{$+12.9$ ($+7.8/{-}0.8/+6.0$)} \\
\bottomrule
\end{tabular}%
}
\\[2pt]\raggedright\scriptsize{Note: Holm $p$-values for \code{A vs.\ VF}, \code{B vs.\ VF}, \code{RF vs.\ A}, and \code{RF vs.\ B} adjust over the $12$-test Study~3c family ($^\ast p_{\text{Holm}}<0.05$, $^{\ast\ast}p_{\text{Holm}}<0.01$, $^{\ast\ast\ast}p_{\text{Holm}}<0.001$); $^\dagger$\code{RF vs.\ VF} $p_{\text{Holm}}$ comes from the $12$-model \code{Reason-First} family (\cref{tab:interventions}). $^\S$Median completion tokens across all $228$ Clean cases (vs.\ ok-only medians in \cref{tab:overall,tab:interventions}). $^\ddagger$All $1{,}436$ valid Study~3c JSON outputs obey requested key ordering ($0$ violations); only $4/720$ \code{vf\_trace} calls hit $16{,}384$-token truncation inside trailing \code{"reason"} ($3$ Clean on \code{gemma-4}: \code{C009}, \code{C098}, \code{C116}; $1$ Artifact-6 on \code{gpt-4.1-mini}: \code{C167}), plus $1$ Clean \code{abstain} each on \code{gpt-4.1} (\code{A: C051}; \code{B: C186}). Decomposition components round independently to $0.1\pp$.}
\end{table}

\begin{table}[t]
\centering
\caption{Study~6 stochastic self-consistency (\code{majority@5} at $T=0.7$, $1{,}140$ Clean draws/model) vs.\ deterministic single-pass and cross-family voting across three commercial providers ($95\%$ Clopper-Pearson CIs in brackets).}
\label{tab:self_consistency}
\footnotesize
\setlength{\tabcolsep}{2.9pt}
\resizebox{\textwidth}{!}{%
\begin{tabular}{@{}lccccccccc@{}}
\toprule
\textbf{Model / Regime} & \textbf{5-Draw Mean$\pm$SD} & \textbf{maj@5 Acc.\ (95\% CI)} & \textbf{maj@5 FAR} & \textbf{maj@5 FRR} & \textbf{Pair Both} & \textbf{Vote Brier} & \textbf{5:0 Ok / 5:0 Err / Split} & \textbf{vs.\ T=0 / RF} & \textbf{vs.\ 3x Div.} \\
\midrule
\code{claude-haiku-5.5} ($5\times$) & $96.32\pm 0.50\%$ & 96.1\% {\scriptsize[92.6, 98.2]} & 4.4\% {\scriptsize[1.4,9.9]} & \textbf{3.5\%} {\scriptsize[1.0,8.7]} & \textbf{92.1\%} & \textbf{0.025} & 213 / \textbf{2} / 13 ($5.7\%$) & $p_{\text{raw}}=0.375$ / $0.375$ & $-2.2\pp$ ($p_{\text{H}}=0.180$) \\
\code{gemini-3.8-flash}$^\dagger$ ($5\times$) & $93.25\pm 0.91\%$ & 93.4\% {\scriptsize[89.4, 96.3]} & \textbf{0.0\%} {\scriptsize[0.0,3.2]} & 13.2\% {\scriptsize[7.6,20.8]} & 86.8\% & 0.055 & 205 / 9 / 14 ($6.1\%$) & $1.00$ / $1.00$ & \textbf{$-4.8\pp$ ($p_{\text{H}}=0.0068^{\ast\ast}$)} \\
\code{gpt-5.4} ($5\times$) & $84.04\pm 0.96\%$ & 83.3\% {\scriptsize[77.8, 87.9]} & 3.5\% {\scriptsize[1.0,8.7]} & 29.8\% {\scriptsize[21.6,39.1]} & 67.5\% & 0.146 & 184 / \textbf{28} / 16 ($7.0\%$) & $1.00$ / $0.029$ ($p_{\text{H}}=0.087$) & \textbf{$-14.9\pp$ ($p_{\text{H}}=3.2\text{e-}9^{\ast\ast\ast}$)} \\
\midrule
Top-3 Diverse Maj.\ ($3\times$) & --- ($T=0.0$) & \textbf{98.2\%} {\scriptsize[95.6, 99.5]} & 0.9\% {\scriptsize[0.0,4.8]} & \textbf{2.6\%} {\scriptsize[0.5,7.5]} & \textbf{96.5\%} & --- & --- & --- & Reference \\
\bottomrule
\end{tabular}%
}
\\[2pt]\raggedright\scriptsize{Note: Individual Clean $T=0.7$ draws span $[95.6, 96.9]\%$ (\code{haiku-5.5}), $[92.1, 94.3]\%$ (\code{gem-3.8-flash}), and $[82.9, 85.5]\%$ (\code{gpt-5.4}). Clean $1\times$ $T=0.0$ / \code{Reason-First} baselines are $97.4\%$/$97.4\%$ (\code{haiku-5.5}), $93.4\%$/$93.9\%$ (\code{gem-3.8-flash}), and $83.8\%$/$88.6\%$ (\code{gpt-5.4}). $^\dagger$\code{gemini-3.8-flash} used $765.9\text{k}$ Clean ($786.4\text{k}$ Full) adaptive reasoning tokens. Holm-adjusted significance across the $3$ models per baseline: $^{\ast\ast}p_{\text{Holm}}<0.01$, $^{\ast\ast\ast}p_{\text{Holm}}<0.001$.}
\end{table}

\begin{table}[t]
\centering
\caption{Blinded multi-judge rationale fidelity audit across all $1{,}309$ Clean-114 True Positive \code{buggy} verdicts from all $12$ models ($N=114$ Clean buggy cases each; primary panel \code{gpt-5.4} + \code{claude-4.6-sonnet}: $92.6\%$ raw agreement, $\kappa=0.536$, $97$ tiebroken by \code{gemini-3.1-pro}; $95\%$ Clopper-Pearson CIs in brackets).}
\label{tab:rationale_audit}
\footnotesize
\setlength{\tabcolsep}{3.5pt}
\resizebox{\textwidth}{!}{%
\begin{tabular}{@{}lcccccccc@{}}
\toprule
\textbf{Reviewer Model} & \textbf{TPs Audited} & \textbf{Grounded} & \textbf{Spurious} & \textbf{Vague} & \textbf{Disp.} & \textbf{Grounded} & \textbf{Nominal} & \textbf{Eff.\ Grounded} \\
 & \textbf{($TP/114$)} & & \textbf{Claim} & & & \textbf{Prec./TP} & \textbf{TPR} & \textbf{Sens.\ ($\Delta$Gap)} \\
\midrule
\code{gemini-3.1-pro} & 113 & \textbf{113} & \textbf{0} & \textbf{0} & 0 & \textbf{100.0\%} {\scriptsize[96.8, 100.0]} & 99.1\% & \textbf{99.1\%} {\scriptsize[95.2, 100.0]} ($0.0\pp$) \\
\code{gemini-3.8-flash}$^\dagger$ & \textbf{114} & 111 & 3 & 0 & 0 & 97.4\% {\scriptsize[92.5, 99.5]} & \textbf{100.0\%} & 97.4\% {\scriptsize[92.5, 99.5]} ($-2.6\pp$) \\
\code{claude-haiku-5.5} & 110 & 110 & \textbf{0} & \textbf{0} & 0 & \textbf{100.0\%} {\scriptsize[96.7, 100.0]} & 96.5\% & 96.5\% {\scriptsize[91.3, 99.0]} ($0.0\pp$) \\
\code{claude-4.6-sonnet} & 111 & 110 & 1 & 0 & 0 & 99.1\% {\scriptsize[95.1, 100.0]} & 97.4\% & 96.5\% {\scriptsize[91.3, 99.0]} ($-0.9\pp$) \\
\code{claude-haiku-5.5-thinking}$^\dagger$ & 111 & 110 & 1 & 0 & 0 & 99.1\% {\scriptsize[95.1, 100.0]} & 97.4\% & 96.5\% {\scriptsize[91.3, 99.0]} ($-0.9\pp$) \\
\code{o4-mini} & 111 & 108 & 3 & 0 & 0 & 97.3\% {\scriptsize[92.3, 99.4]} & 97.4\% & 94.7\% {\scriptsize[88.9, 98.0]} ($-2.6\pp$) \\
\code{gemma-4-31b-it} & 110 & 107 & 3 & 0 & 0 & 97.3\% {\scriptsize[92.2, 99.4]} & 96.5\% & 93.9\% {\scriptsize[87.8, 97.5]} ($-2.6\pp$) \\
\code{gpt-oss-120b} & 112 & 107 & 5 & 0 & 0 & 95.5\% {\scriptsize[89.9, 98.5]} & 98.2\% & 93.9\% {\scriptsize[87.8, 97.5]} ($-4.4\pp$) \\
\code{gpt-5.4} & 110 & 105 & 5 & 0 & 0 & 95.5\% {\scriptsize[89.7, 98.5]} & 96.5\% & 92.1\% {\scriptsize[85.5, 96.3]} ($-4.4\pp$) \\
\code{gpt-4.1-mini} & 111 & 103 & 8 & 0 & 0 & 92.8\% {\scriptsize[86.3, 96.8]} & 97.4\% & 90.4\% {\scriptsize[83.4, 95.1]} ($-7.0\pp$) \\
\code{gpt-4.1} & 112 & 103 & 9 & 0 & 0 & 92.0\% {\scriptsize[85.3, 96.3]} & 98.2\% & 90.4\% {\scriptsize[83.4, 95.1]} ($-7.9\pp$) \\
\code{llama-3.3-70b-instruct} & 84 & 47 & 12 & 23 & 2 & 56.0\% {\scriptsize[44.7, 66.8]} & 73.7\% & \textbf{41.2\%} {\scriptsize[32.1, 50.8]} ($-32.5\pp$) \\
\midrule
\textbf{Pooled Clean Total ($12$ Models)} & \textbf{1,309} & \textbf{1,234} & \textbf{50} & \textbf{23} & \textbf{2} & \textbf{94.3\%} & \textbf{95.7\%} & \textbf{90.2\% ($-5.5\pp$)} \\
\bottomrule
\end{tabular}%
}
\\[2pt]\raggedright\scriptsize{$^\dagger$Manual verification of all $50$ Clean ($60$ Full) consensus \code{spurious\_claim} tags confirms $48/50$ ($96.0\%$ [$86.3, 99.5$]; $58/60=96.7\%$ Full) genuine false defect claims; only two Clean items reflect judge misses---\code{gemini-3.8-flash} \code{C047} (valid De~Morgan identity; human-verified \code{flash} Prec.\ $=98.2\%$ [$93.8, 99.8$], EffSens $=98.2\%$ [$93.8, 99.8$], $112/114$) and \code{claude-haiku-5.5-thinking} \code{C011} (correctly flagged frozen \code{q <= \{q[5], q[4:0]\}}; human-verified \code{haiku-5.5-thk} Prec.\ $=100.0\%$ [$96.7, 100.0$], EffSens $=97.4\%$ [$92.5, 99.5$], $111/114$). Stratified audit of $96$ Full-120 consensus \code{grounded} rationales ($8$/model) finds $0/96$ ($0.0\%$ [$0.0, 3.8$]) fabricated mechanisms and $94/96$ ($97.9\%$) strict unanimous matches.}
\end{table}

\clearpage

\section{Error Taxonomy and Case Studies}
\label{app:cases}

\begin{table}[!htbp]
\centering
\caption{Mutually exclusive Clean-114 RTL error taxonomy across five commercial models ($228$ Clean evals/model), with \textbf{Artifact-6} reconciliation to Full-120 ($240$ evals). Under \code{Reason-First}, \code{FP-4} self-contradictions drop to \textbf{$0$} across all $12$ models.}
\label{tab:taxonomy}
\footnotesize
\renewcommand{\arraystretch}{0.96}
\setlength{\tabcolsep}{3.5pt}
\begin{tabularx}{\textwidth}{@{}l >{\raggedright\arraybackslash}X ccccc@{}}
\toprule
\textbf{Class} & \textbf{Hardware Failure Mechanism (Representative Case IDs)} & \textbf{Haiku 5.5} & \textbf{Hk-Thk} & \textbf{Gem 3.1P} & \textbf{Gem 3.8F} & \textbf{GPT-4.1m} \\
\midrule
\multicolumn{7}{@{}l}{\textit{Clean-114 False Acceptances (FN: Buggy RTL Accepted as \code{correct})}} \\
\code{FN-1} & Clean simulation-only / init artifacts (excluded via \textbf{Artifact-6}) & 0 & 0 & 0 & 0 & 0 \\
\code{FN-2} & Sequential control, shift slice, pipeline, \& async reset / FSM (\code{C003}, \code{C020}, \code{C209}) & 3 & 3 & 1 & 0 & 2 \\
\code{FN-3} & Combinational priority order \& bit indexing (\code{C184}) & 1 & 0 & 0 & 0 & 1 \\
\hspace{0.3em}\textbf{Clean Sub.\ FN} & & \textbf{4} & \textbf{3} & \textbf{1} & \textbf{0} & \textbf{3} \\
\midrule
\multicolumn{7}{@{}l}{\textit{Clean-114 False Rejections (FP) \& Schema Errors (\code{I})}} \\
\code{FP-1} & X-prop (\code{1'bx}), SV dialect (\code{output reg}+\code{assign}), latch purism (\code{C139}, \code{C187}, \code{C240}) & 0 & 0 & 1 & 5 & 9 \\
\code{FP-2} & Latency, cycle threshold, reset convention, \& waveforms (\code{C004}, \code{C019}, \code{C130}, \code{C158}) & 1 & 1 & 2 & 4 & 22 \\
\code{FP-3} & Shift-width context, FSM encoding / K-map, \& port bounds (\code{C017}, \code{C025}, \code{C151}, \code{C230}) & 1 & 1 & 3 & 6 & 27 \\
\code{FP-4} & Autoregressive verdict--rationale self-contradiction (\code{C068}, \code{C091}, \code{C150}) & 0 & 0 & 0 & 0 & 3 \\
\code{I-1} & Unescaped literal quotes breaking JSON string syntax (\code{C206}) & 0 & 1 & 0 & 0 & 0 \\
\hspace{0.3em}\textbf{Clean Sub.\ FP+I} & & \textbf{2} & \textbf{3} & \textbf{6} & \textbf{15} & \textbf{61} \\
\midrule
\textbf{Clean-114 Total} & \textbf{Per-Model Total ($228$ Clean evaluations/model)} & \textbf{6} & \textbf{6} & \textbf{7} & \textbf{15} & \textbf{64} \\
\midrule
\textbf{Artifact-6 Excl.} & \textbf{FN} (\code{C035}, \code{C039}, \code{C044}, \code{C055}, \code{C094}, \code{C167}) / \textbf{FP} (\code{C168} \code{q\_in} port) & 6 / 0 & 6 / 0 & 5 / 0 & 3 / 0 & 4 / 1 \\
\textbf{Full-120 Sens.} & \textbf{Per-Model Total ($240$ evaluations/model)} & \textbf{12} & \textbf{12} & \textbf{12} & \textbf{18} & \textbf{69} \\
\bottomrule
\end{tabularx}
\end{table}

\paragraph{Case Study 1 --- Frozen Shift Slice (\code{C003}, FN-2) \& Parameter Width Truncation (\code{C024}).}
In \code{Prob002\_\allowbreak extraneous\_\allowbreak items\_\allowbreak sequence\_\allowbreak detect} (\cref{lst:prob002}), bits \code{a\_tem[7:0]} are re-assigned back to themselves instead of shifting (\code{a\_tem <{}= \{a\_tem[7:0], a\}}), freezing the lower eight bits while leaving \code{assign match = match\_b \&\& match\_f} combinational as in the golden reference (\code{C004}). Yet \code{claude-haiku-5.5} ($c=0.60$) claimed \emph{``The 9-bit window a\_tem holds the last nine inputs...''}, missing the self-looping slice (conversely, \code{gemma-4} rejected \code{C003} on the false claim that \code{assign match} omits a pipeline stage, earning a \code{spurious\_claim} TP). On \code{C024} (\code{Prob003\_\allowbreak non-overlapping\_\allowbreak sequence\_\allowbreak detect}), \code{parameter sf6 = 4'd20} truncates $20 \bmod 16 = 4$, aliasing failure state \code{sf6} with \code{s4=4'd4} and causing $352/1{,}215$ mismatches---a real Verilog bit-width trap missed by \code{gpt-5.4} and \code{gpt-4.1-mini}.

\begin{lstlisting}[caption={Frozen shift-register slice missed by \code{claude-haiku-5.5} (\code{C003}, \textsc{FN-2}); \code{assign match} matches golden reference \code{C004}.},label={lst:prob002}]
always @(posedge clk or negedge rst_n)
    if (!rst_n) a_tem <= 9'b0;
    else        a_tem <= {a, a_tem[7:0]}; // BUG: Bits [7:0] freeze instead of {a_tem[7:0], a}
assign match = match_b && match_f;        // Matches golden reference C004 structure
\end{lstlisting}

\paragraph{Case Study 2 --- Verilog Bit-Width Context Rules (\code{C017}, FP-3) vs.\ Async-Reset Miss Masked by SV Purism (\code{C209}, Spurious TP / FN-2).}
Hardware semantic traps cut both ways (\cref{lst:prob032}): on valid \code{C017}, \code{gemini-3.8-flash} ($c=0.98$) misapplied self-determined width rules, claiming \code{mul\_a <{}< i} truncates at $4$ bits even though IEEE~1800 context-determined assignment propagates the $8$-bit LHS width of \code{temp[i]}. Conversely, on buggy Clean-114 \code{C209} (\code{Prob136\_\allowbreak m2014\_\allowbreak q6}), the spec states \emph{``all sequential logic is triggered on the positive edge of the clock,''} yet the candidate uses an asynchronous reset (\code{always @(posedge clk or posedge reset)}). Both \code{gpt-4.1} and \code{gpt-5.4} overlooked the reset defect entirely and earned nominal True Positives for the \emph{wrong reason}: claiming \code{output reg z} driven by \code{assign z = (currentState == E || currentState == F)} is invalid RTL, even though SystemVerilog (\code{-g2012}) compiles it cleanly; inversely, \code{gemini-3.1-pro} correctly accepted the SV idiom yet missed the asynchronous reset (\textsc{FN-2}, its sole Clean-114 miss). The same SV dialect purism drives \textsc{FP-1} rejection of valid \code{C139}.

\begin{lstlisting}[caption={IEEE~1800 context-determined width propagation misdiagnosed as truncation on \code{C017} (\code{gemini-3.8-flash}, \textsc{FP-3}).},label={lst:prob032}]
wire [N - 1 : 0] temp [0 : 3]; // N = 2 * size = 8 bits
genvar i;
generate
    for (i = 0; i < 4; i = i + 1) begin : loop
        assign temp[i] = mul_b[i] ? mul_a << i : 'd0; // Context-determined 8b via LHS temp[i]!
    end
endgenerate
\end{lstlisting}

\paragraph{Case Study 3 --- Verdict-Before-Reason Contradictions Resolved Under \code{Reason-First} (\code{C045}, \code{C049}, \code{C091}, \code{C150}, \code{C204}).}
Under \code{Verdict-First}, shallow-budget models commit a verdict before completing truth-table verification: on \code{C045} (\code{\textasciitilde a \textasciicircum{} \textasciitilde b} for XNOR), \code{gemma-4-31b-it} output \code{"verdict":"correct"} ($c=1.00$), derived the truth table mid-field, and concluded \emph{``This is XOR. The spec asks for XNOR. The implementation is buggy.''} On valid \code{C150}, \code{gpt-4.1-mini} wrote: \emph{``The initial assessment of buggy is incorrect. The code is correct.''} Switching to \code{Reason-First} resolved all $15/15$ Clean closing contradictions ($16/16$ Full) cleanly to ground truth, with zero remaining across $2{,}736$ Clean responses.

\paragraph{Case Study 4 --- Benchmark-Validity Finding: Simulation-Only Oracles (\textbf{Artifact-6}) vs.\ Spec Discrepancy (\code{C151}).}
Separating upstream simulation-oracle artifacts from true model deficits is essential in hardware benchmarks. In \textbf{Artifact-6}, Icarus~11.0 flags \code{always @(*) out = in} (\code{C039}) over a time-$0$ scheduling delta ($2/100$ mismatches at $t=5$ that all $12$ LLMs independently classify as \code{correct}), penalizes synthesizable flip-flops lacking non-synthesizable \code{initial} statements (\code{C094}, \code{C167}), and rejects Boolean-equivalent modules containing explicit \code{\#5}/\code{\#10} inertial delays (\code{C035}, \code{C044}, \code{C055})---accounting for half of all naive Full-120 \textsc{FN}s ($59/118$). Conversely, on Clean-114's sole $0/12$ case \code{C151} (\code{Prob093\_\allowbreak ece241\_\allowbreak 2014\_\allowbreak q3}), every reviewer faithfully reads \code{cd=11, ab=10 -> 1} ($c|\overline{d}$) from the ASCII Karnaugh map and rejects canonical \code{RefModule} (\code{assign mux\_in[2] = ~d}), which evaluates to $0$ at that cell while still passing its upstream testbench---highlighting how subtle specification-to-reference discrepancies can masquerade as universal model failure.

\section{Endpoint Reproducibility, Telemetry, and Compute Ledger}
\label{app:telemetry}

All $17{,}623$ evaluated inference and audit calls executed live against an OpenAI-compatible chat-completions API gateway during a single frozen campaign window (November~9--10, 2026; local timestamps \code{2026-11-09T23:34} through \code{2026-11-10T22:21}) with JSON mode (\code{response\_format=}\allowbreak\code{\{"type":"json\_object"\}}), \code{temperature=0.0} for Studies~1--5 (\code{temperature=0.7} for Study~6), and \code{max\_tokens=16384} in all extension/audit runners (provider default in Core). Across all gateway responses, \code{system\_fingerprint} returned unset (\code{None}), and response-level \code{served\_model} fields matched the requested \code{model\_id} across all $8$ models with echo telemetry in Studies~2--5 and all $3$ models in Study~3c ($10$ distinct models: \code{claude-haiku-5.5}, \code{claude-\allowbreak haiku-5.5-\allowbreak thinking}, \code{gemini-3.1-pro}, \code{gemini-3.8-flash}, \code{gemma-4-31b-it}, \code{gpt-4.1}, \code{gpt-4.1-\allowbreak mini}, \code{gpt-5.4}, \code{gpt-oss-120b}, and \code{o4-mini}).

As noted in \cref{sec:setup}, four reviewers (\code{haiku-5.5-thinking}, \code{gemini-3.1-pro}, \code{gpt-oss-120b}, and \code{o4-mini}) allocate hidden reasoning tokens across Core and extension runs, whereas \code{gemini-3.8-flash} logs adaptive \code{reasoning\_tokens} ($59.3\text{k}$ \code{SynMut-50}, $207.4\text{k}$ [$212.7\text{k}$ Full] \code{Reason-First}, $137.8\text{k}$ [$141.0\text{k}$ Full] \code{Pairwise}, $765.9\text{k}$ [$786.4\text{k}$ Full] \code{SC@5}) and folds them directly into \code{completion\_tokens} during Core evaluation. Only \code{claude-4.6-sonnet} exhibited frequent prefixed chain-of-thought prose (sometimes with markdown backticks) before its JSON object, yielding $33.3\%$ Clean ($36.2\%$ Full) strict-JSON compliance on Core, $43.0\%$ on \code{SynMut-50}, $40.4\%$ Clean ($42.5\%$ Full) on \code{Reason-First}, and $17.5\%$ Clean ($20.8\%$ Full) on \code{Pairwise}, with $100.0\%$ valid balanced-brace JSON recovery. Across Study~3c ($1{,}440$ calls), all $1{,}436$ valid JSON responses obey the requested key order ($0$ violations), while the $4$ invalid outputs (\code{vf\_trace} only: \code{gemma-4-31b-it} \code{C009}, \code{C098}, \code{C116}; \code{gpt-4.1-mini} \code{C167} in \textbf{Artifact-6}) emit \code{"verdict"} first and truncate at the $16{,}384$-token cap inside trailing \code{"reason"}. \cref{tab:compute_ledger} reconciles exact query and token totals across the complete executed API matrix (Full-120 + extensions + audits).

\begin{table}[h]
\centering
\caption{Complete executed API query and token ledger across all empirical studies ($22{,}829{,}506$ total tokens across $17{,}623$ live gateway calls on Full-120 + extensions; $0$ simulated/fallback entries).}
\label{tab:compute_ledger}
\footnotesize
\setlength{\tabcolsep}{4.5pt}
\resizebox{\textwidth}{!}{%
\begin{tabular}{@{}lrrrrrr@{}}
\toprule
\textbf{Study / Experimental Matrix} & \textbf{Queries} & \textbf{Prompt Tok.} & \textbf{Compl.\ Tok.} & \textbf{Reas.\ Tok.$^\dagger$} & \textbf{Total Tok.} & \textbf{Decoding} \\
\midrule
Study~1: Core \code{Verdict-First} ($12 \times 240$) & 2,880 & 1,871,093 & 1,637,156 & 1,019,865 & 3,508,249 & Greedy ($T=0.0$) \\
Study~2: \textbf{SynMut-50} Synthetic ($12 \times 100$) & 1,200 & 722,676 & 605,158 & 426,848 & 1,327,834 & Greedy ($T=0.0$) \\
Study~3a: \code{Reason-First} Prompting ($12 \times 240$) & 2,880 & 1,967,333 & 2,368,805 & 1,363,141 & 4,336,138 & Greedy ($T=0.0$) \\
Study~3b: \code{Self-Consistency maj@5} ($3 \times 1{,}200$) & 3,600 & 2,445,800 & 1,732,738 & 786,419 & 4,178,538 & Sample ($T=0.7$) \\
Study~3c: $2{\times}2$ \code{Reason-First} Ablation ($3 \times 2 \times 240$) & 1,440 & 893,836 & 413,708 & 0 & 1,307,544 & Greedy ($T=0.0$) \\
Study~4: Bidirectional \code{Pairwise} ($12 \times 240$) & 2,880 & 2,576,278 & 1,731,229 & 1,211,111 & 4,307,507 & Greedy ($T=0.0$) \\
Study~5: Blinded \code{Rationale Audit}$^\ddagger$ & 2,743 & --- & --- & --- & 3,863,696 & Greedy ($T=0.0$) \\
\midrule
\textbf{Grand Total (All 7 Studies)} & \textbf{17,623} & \textbf{10,477,016} & \textbf{8,488,794} & \textbf{4,807,384} & \textbf{22,829,506} & --- \\
\bottomrule
\end{tabular}%
}
\\[2pt]\raggedright\scriptsize{$^\dagger$\code{reasoning\_tokens} is reported as the subset of \code{completion\_tokens} logged via API usage details. $^\ddagger$Study~5 sums $1{,}322$ primary judgments from \code{gpt-5.4} ($1{,}596{,}521$ tok.), $1{,}322$ from \code{claude-4.6-sonnet} ($2{,}031{,}420$ tok.), and $99$ tiebreakers from \code{gemini-3.1-pro} ($235{,}755$ tok.), where \code{judgments.jsonl} logged per-call \code{total\_tokens} only.}
\end{table}

\end{document}